\documentclass[twocolumn]{aastex63}
\usepackage{lineno}
\usepackage{amsmath}
\usepackage{multirow}

\newcommand{\lya}{Ly$\alpha$}

\shorttitle{QLF at $3.5<z<5.0$}
\shortauthors{Pan et al.}

\begin{document}
\title{Quasar UV Luminosity Function at $3.5<z<5.0$ from SDSS Deep Imaging Data}

\author[0000-0003-0230-6436]{Zhiwei Pan}
\altaffiliation{panzhiwei@pku.edu.cn}
\affiliation{Kavli Institute for Astronomy and Astrophysics, Peking University, Beijing 100871, China}
\affiliation{Department of Astronomy, School of Physics, Peking University, Beijing 100871, China}

\author[0000-0003-4176-6486]{Linhua Jiang}
\altaffiliation{jiangKIAA@pku.edu.cn}
\affiliation{Kavli Institute for Astronomy and Astrophysics, Peking University, Beijing 100871, China}
\affiliation{Department of Astronomy, School of Physics, Peking University, Beijing 100871, China}

\author[0000-0003-3310-0131]{Xiaohui Fan}
\affiliation{Steward Observatory, University of Arizona, 933 North Cherry Avenue, Tucson, AZ 85721, USA}

\author[0000-0001-5364-8941]{Jin Wu}
\affiliation{Kavli Institute for Astronomy and Astrophysics, Peking University, Beijing 100871, China}

\author[0000-0001-5287-4242]{Jinyi Yang}
\affiliation{Steward Observatory, University of Arizona, 933 North Cherry Avenue, Tucson, AZ 85721, USA}

\begin{abstract}
We present a well-designed sample of more than 1000 type 1 quasars at $3.5<z<5$ and derive UV quasar luminosity functions (QLFs) in this redshift range. These quasars were selected using the Sloan Digital Sky Survey (SDSS) imaging data in SDSS Stripe 82 and overlap regions with repeat imaging observations. They are about one magnitude fainter than those found using the SDSS single-epoch data. The spectroscopic observations were conducted by the SDSS-III Baryon Oscillation Spectroscopic Survey (BOSS) as one of the BOSS ancillary programs. This quasar sample reaches $i\sim21.5$ mag and bridges previous samples from brighter surveys and deeper surveys. We use a $1/V_\mathrm{a}$ method to derive binned QLFs at $3.6<z<4.0$, $4.0<z<4.5$, and $4.5<z<4.9$, and use a double-power law model to parameterize the QLFs. We also combine our data with those in the literature to better constrain the QLFs in the context of a much wider luminosity baseline. We find that the faint-end and bright-end slopes of the QLFs in this redshift range are around $-1.7$ and $-3.7$, respectively, with uncertainties from 0.2{--}0.3 to $>0.5$. The evolution of the QLFs from $z\sim5$ to $3.5$ can be described by a pure density evolution model ($\propto10^{kz}$) and the parameter $k$ is similar to that at $5<z<7$, suggesting a nearly uniform evolution of the quasar density at $z=3.5-7$.
\end{abstract}


\section{Introduction}
Quasars were first discovered in the radio band \citep{1963Natur.197.1040S} and were soon recognized as luminous extra-galactic sources in multiple bands from radio to X-ray. The tremendous energy of quasars originates from the accretion of their central super-massive black holes (SMBHs). Due to their high luminosities, quasars are powerful tools to probe the distant universe. They are often used to study SMBHs, their host galaxies, the intergalactic medium, etc. A number of surveys have been carried out to search for quasars in the past twenty years, such as the 2dF QSO Redshift Survey \citep{2000MNRAS.317.1014B}, the 6dF QSO Redshift Survey \citep{2004MNRAS.349.1397C}, the Sloan Digital Sky Survey (SDSS) Quasar Survey \citep{Richards_2006}, and the SkyMapper Southern Survey \\citep{2020MNRAS.491.1970W,2021arXiv210512215O}. The total number of known quasars  has amounted to roughly one million \citep{flesch2021million} and SDSS contributed the majority of the known quasars \citep{2020ApJS..250....8L}. The most distant quasar known to date is at $z=7.642$ \citep{2021ApJ...907L...1W}.

Quasar luminosity function (QLF) has been widely used to measure how the spatial density of quasars evolves with luminosity and redshift. It has also been used to constrain the quasar contribution to the cosmic X-ray and infrared background \citep[e.g.,][]{2001ARA&A..39..249H,Hopkins_2007,10.1093/mnras/staa1381} and the contribution of quasar UV photons to the cosmic H and He reionization \citep[e.g.,][]{2011ApJ...733L..24W,2016ApJ...833..222J}. The QLF at $z<3.5$ in the UV/optical band has been well studied. A pure luminosity evolution (PLE) model with a double power-law shape can efficiently describe the QLF at $z=0-2$ \citep[e.g.,][]{1988MNRAS.235..935B,2000MNRAS.317.1014B,2004MNRAS.349.1397C}, suggesting that the characteristic luminosity in the QLF evolves with redshift while the faint- and bright-end slopes remain unchanged in this redshift range. The PLE scenario is not enough to describe the QLF at higher redshifts. A luminosity evolution and density evolution (LEDE) model was proposed to fit the QLF at $z\sim2-3.5$ \citep[e.g.,][]{Ross_2013}.

At $z>3.5$, the bright-end QLF has been measured reasonably well, but the faint end has not been well determined. A full QLF fit usually relies on the combination of large-scale surveys (e.g., SDSS) and small pencil-beam surveys \citep[e.g.,][]{Glikman_2010,Glikman_2011}. Using early SDSS data, \citet{2001AJ....121...54F} studied 39 luminous quasars and suggested that the bright-end shape of the QLF evolves with redshift at $z>3$. \citet{Glikman_2010,Glikman_2011} studied the faint end of the QLF at $z\sim4$ using 23 quasars and found a shallow slope $\alpha=-1.6$ that is consistent with previous studies. For QLFs at $z\sim5$,  \citet{2013ApJ...768..105M} constructed a well-defined sample of 71 quasars from SDSS and measured the QLF at $4.7<z<5.1$.  \citet{2016ApJ...829...33Y} extended the bright end of the QLF at $z\sim5$. Table \ref{table:qlf_overview} lists some recent studies of UV/optical QLFs that cover a redshift range of $z\sim3-5$. Recently, QLFs at $z\sim6-7$ are also being established \citep[e.g.,][]{2016ApJ...833..222J,2018ApJ...869..150M,Wang_2019}.

As seen above, significant progress has been made in determining QLFs in different redshift and luminosity ranges. However, the evolution of the quasar population in a wide redshift and luminosity range has not been well characterized. Some studies have tried to analyze such an evolution based on the combination of different quasar samples from the literature \citep{2017MNRAS.466.1160M,2019MNRAS.488.1035K,10.1093/mnras/staa1381,2021arXiv210306265K}. For example, \citet{2021arXiv210306265K} selected and combined some binned QLFs at $z\sim2.4,3.9,5.0,6.1$ from the literature. They found that a pure density evolution (PDE) model is enough to describe the QLFs at $2<z<6$. In such studies, one has to assume that there are no systematic effects among the individual measurements of QLFs, which is not always the case.

In this paper, we present more than one thousand quasars identified by SDSS. The majority of these quasars form a well-designed sample at $3.5<z<5$ that allows us to derive reliable QLFs in this redshift range. This sample is roughly one magnitude fainter than the SDSS main quasar sample \citep{Richards_2006}. The layout of the paper is as follows. In Section \ref{sec:selection}, we introduce the target selection and spectroscopic observations of our quasar candidates. In Section \ref{sec:results}, we present our quasar sample, calculate its area coverage, estimate sample incompleteness, and derive QLFs . In Section \ref{sec:dis}, we compare our result with previous studies and discuss the evolution of the QLF at high redshift. We summarize the paper in Section \ref{sec:sum}. Throughout this paper, we use PSF (point spread function) magnitudes, and magnitudes are expressed in the AB system (i.e., SDSS magnitudes are converted to AB magnitudes). We adopt a $\Lambda$-dominated flat cosmology with $H_0=70$ km s$^{-1}$ Mpc$^{-1}$, $\Omega_{m}=0.3$, and $\Omega_{\Lambda}=0.7$.

\begin{deluxetable*}{cccccc}
\tablecaption{Selected Studies of Optical QLFs} \label{table:qlf_overview}
\tablewidth{0pt}
\tablehead{Survey & Area (deg$^2$) & $\mathrm{N_Q}\tablenotemark{\scriptsize$1$}$   & Magnitude Range                           & Redshift Range               & Reference}
\startdata   
SDSS                                            & 182            & 39               & $i\leqslant20$                            & $3.6<z<5.0$                  & \citet{2001AJ....122.2833F} \\
COMBO-17                                        & 0.8            & 192              & $R<24$                                    & $1.2<z<4.8$                  & \citet{2003AA...408..499W} \\
SDSS DR3                                        & 1622           & 15,343           & $i\leqslant19.1$ and $i\leqslant20.2$     & $0.3<z<5.0$                  & \citet{Richards_2006} \\
GOODS+SDSS                                      & 0.1+4200       & 13+656           & $22.25<z_{850}<25.25$                     & $3.5<z<5.2$                  & \citet{2007AA...461...39F} \\
VVDS                                            & 0.62           & 130              & $17.5<I_{\mathrm{AB}}<24.0$               & $0<z<5$                      & \citet{2007AA...472..443B} \\
COSMOS                                          & 1.64           & 8                & $22<i'<24$                                & $3.7<z<4.7$                  & \citet{2011ApJ...728L..25I} \\
NDWFS+DLS                                       & 4              & 24               & $R\leqslant24$                            & $3.7<z<5.1$                  & \citet{Glikman_2011} \\
COSMOS                                          & 1.64           & 155              & $16\leqslant I_{\mathrm{AB}} \leqslant25$ & $3<z<5$                      & \citet{2012ApJ...755..169M} \\
SDSS DR7                                        & 6248           & 57,959           & $i\leqslant19.1$ and $i\leqslant20.2$     & $0.3<z<5.0$                  & \citet{2012ApJ...746..169S} \\
BOSS+MMT                                        & 14.5+3.92      & 1877             & $g\lesssim23$                             & $0.7<z<4.0$                  & \citet{2013AA...551A..29P} \\
BOSS+MMT                                        & 235            & 71               & $i<22$                                    & $4.7<z<5.1$                  & \citet{2013ApJ...768..105M} \\
SDSS+WISE                                       & 14,555         & 99               & $z\leqslant19.5$                                   & $4.7<z<5.4$                  & \citet{2016ApJ...829...33Y} \\
CFHTLS                                          & 105+18.5       & 37               & $i_{\mathrm{AB}}<23.7$                    & $4.7<z<5.4$                  & \citet{McGreer_2018} \\
COSMOS                                          & 1.73           & 16               & $i_{\mathrm{AB}}<23.0$                    & $3.6<z<4.2$                  & \citet{Boutsia_2018} \\
ELQS                                            & 11,838.5       & 407              & $m_i\leqslant18.0$                        & $2.8<z<4.5$                  & \citet{Schindler_2019} \\ 
IMS                                             & 85             & 49               & $i'<23$                                   & $4.7<z<5.4$                  & \citet{Kim_2020} \\
QUBRICS                                         & 12,400         & 58               & $i_{\mathrm{psf}}\leqslant18$             & $3.6<z<4.2$                  & \citet{2021arXiv210310446B} \\
BOSS                                            & $\sim1500$     & 1198             & $19.0<i<21.5$                             & $3.6<z<4.9$                  & This paper            
\enddata 
\tablenotetext{$\scriptsize1$}{$\mathrm{N_Q}$ is the number of spectroscopically confirmed quasars.}
\end{deluxetable*} 

\section{Target selection and observations}{} \label{sec:selection}

In this section, we will briefly introduce the SDSS imaging survey, and then present the details of the quasar candidate selection from the SDSS imaging data. Our program was one of the SDSS-III Baryon Oscillation Spectroscopic Survey (BOSS) ancillary programs, so in the end of the section, we will provide a summary of the BOSS spectroscopic observations of the quasar candidates.

\subsection{The SDSS Imaging Survey}

The SDSS is an imaging and spectroscopic survey using a dedicated wide-field 2.5 m telescope \citep{2006AJ....131.2332G} with five broad bands, $ugriz$, at Apache Point Observatory. An SDSS imaging run consists of six parallel scanlines, and two interleaving runs slightly overlap, leading to duplicate
observations in a small area. The imaging survey was along great circles and had two common poles, so regions near the survey poles overlap substantially. In addition, if a run (or part of a run) did not satisfy the SDSS quality criteria, the relevant region was re-observed, yielding duplicate observations in this region. Due to the above survey strategy and geometry, SDSS has a large amount of duplicate observations (referred to as overlap regions in this paper). The total area of the overlap regions is more than one-third of the SDSS footprint. Detailed information about these overlap regions can be found in \citet{2015AJ....149..188J,2016ApJ...833..222J}. We selected quasars in part of the overlap regions in this paper. These overlap regions provide a unique dataset that allows us to select quasars fainter than those found from the SDSS single-epoch data.

\begin{figure*}
\epsscale{0.8}
\plotone{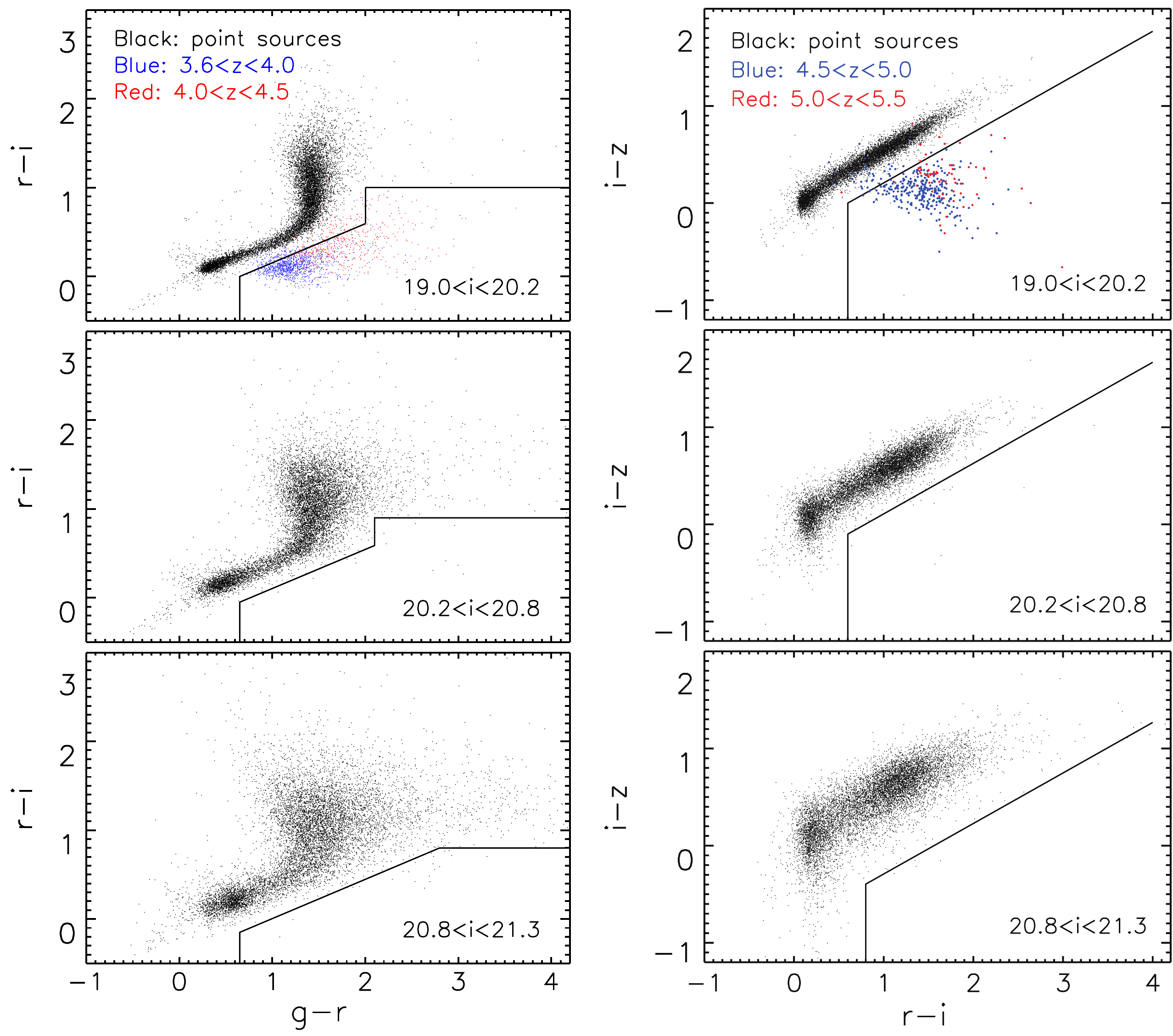}
\caption{The $r-i$ versus $g-r$ (left) and $i-z$ versus $r-i$ (right) color-color diagrams for the illustration of our quasar candidate selection. The black dots are randomly selected point sources that define stellar loci. In the top panels, the blue and red dots represent quasars from the SDSS DR5 quasar catalogue \citep{2007AJ....134..102S} at two redshifts. They are randomly selected with $i<20.2$ mag. The black lines indicate our quasar selection criteria. These criteria are slightly different in different magnitude ranges.  
\label{fig:selection}}
\end{figure*}

In addition to the single-epoch main survey, SDSS conducted a deep imaging survey of $\sim300\ \mathrm{deg}^2$ (Stripe 82) on the Celestial Equator in the SGC \citep{2014ApJ...794..120A,2014ApJS..213...12J}. Stripe 82 roughly spans $20^\mathrm{h}<\mathrm{R.A.}<4^\mathrm{h}$ and $-1\fdg26<\mathrm{Decl.}<1\fdg26$, and was scanned around $70-90$ times. The combined data are 1.5--2 mag deeper than the single-epoch data.

\subsection{Target Selection}
There are a variety of quasar selection methods using optical imaging data. Early searches of type 1 quasars largely rely on point-like morphology and blue UV continuum colors. This color selection is efficient for quasars at relatively low redshift, as quasars and stars have different loci in color-color diagrams \citep{1999AJ....117.2528F}. Later, more methods and more sophisticated techniques were developed. For example, transfer learning has been a useful tool \citep{2021arXiv210209770F} for sky regions with large dust extinction and large contamination like the Galactic plane. Other methods including the likelihood approach \citep{Kirkpatrick_2011}, the Neutral Network approach \citep{2010A&A...523A..14Y}, and the Extreme Deconvolution \citep{Bovy_2011} have been applied to recent quasar surveys such as the SDSS-III BOSS quasar survey \citep{2012ApJS..199....3R}.

Our goal was to select quasars at $3.6<z<5.5$. Searches of higher-redshift quasars in SDSS have been carried out by other programs \citep[e.g.,][]{2006AJ....131.1203F,2016ApJ...833..222J}. We chose to use the traditional color-color diagrams to select our targets. Specifically, we selected quasars at $3.6<z<4.5$ and $4.5<z<5.5$ using the $gri$ and $riz$ colors, respectively (see Figure \ref{fig:selection} and Table \ref{table:total}). At $z>3.6$ ($z>4.5$), the \lya\ emission line enters the $r$ ($i$) band, and the Lyman forest absorption makes quasars much fainter in bluer bands. Therefore, the color-color diagrams are efficient for the selection of quasars in these two redshift ranges \citep{1999AJ....118....1F,2002AJ....123.2945R}. 

Our targets (like other targets for the SDSS BOSS survey) were selected from the SDSS Data Release 7 (DR7) imaging data. We first present our quasar selection in the overlap regions. We use `1' to denote the SDSS primary detection and `2' to denote the SDSS secondary detection. For example, $i_1$ ($i_2$) is the $i$-band magnitude for the primary (secondary) detection. SDSS primary detections generally have slightly higher signal-to-noise ratios than secondary detections. The selection procedure consists of two major steps. In the first major step, we retrieved a preliminary candidate list from the SDSS Query CasJobs online server. We searched the following area at high Galactic latitude: $\rm 100\degr<R.A.<300\degr$, $\rm Decl.>-5\degr$, and Galactic latitude $b>40\degr$. The object type of both primary and secondary detections is `star', i.e., they were classified as point sources. Although distant quasars are point-like objects in ground-based images, faint point sources can be misclassified as extended objects. We will correct this effect in Section \ref{sec:results}. The positional separation between a primary detection and its secondary detection was required to be smaller than $0\farcs5$, which ensures that they are the same object. We excluded objects with the SDSS processing flags `BRIGHT', `EDGE', `SATUR', and `BLENDED'. We then imposed initial color cuts to reduce the number of objects in the preliminary target list. The initial cuts are similar to (but much looser than) the final color cuts addressed in Appendix \ref{App:sel}. We do not expect to lose real quasars in this step. In addition, objects with previous spectroscopic observations were excluded using `specObjID'$=0$ (meaning no spectroscopic observations). This was to reduce the number of targets for follow-up spectroscopy. 

After we obtained the preliminary list of targets, we combined the primary and secondary detections. For each object, we first converted its primary and secondary magnitudes to flux, and calculated weighted mean flux. Errors were added in quadrature. We then converted the combined flux and errors to AB magnitudes and errors. For example, $i$ ($i_{\rm{err}}$) denote the combined $i$-band magnitude (error). The combined magnitudes and errors will be used in the following color selection.

\begin{deluxetable*}{cccccccc}
\tablecaption{Summary of the Quasar Samples} \label{table:total}
\tablewidth{0pt}
\tablehead{& \multicolumn{3}{c}{New Sample} & \multicolumn{3}{c}{Archival sample}  & Total sample  \\
& $gri$  & $riz$  & all   & $gri$  &  $riz$   &all   &  all  }
\startdata   
Overlap regions & 589 & 63 & 652  & 282  &  40  &322  &  974 \\
Stripe 82 &  176 & 24 & 200  & 21 & 3 & 24  &224  \\
All &  &  & 852  &  &  & 346 &  1198\\
\enddata
\end{deluxetable*} 

Our color selection criteria were primarily based on the criteria for SDSS I and II from \citet{2002AJ....123.2945R}. The selection of candidates at $3.6<z<4.5$ ($gri$ candidates) was based on the $r-i$ versus $g-r$ diagram (the left column of Figure \ref{fig:selection}). In Figure \ref{fig:selection}, the black dots represent randomly selected point sources from the SDSS Query CasJobs online server, the blue and red dots represent a sample of randomly selected quasars from SDSS DR5 \citep{2007AJ....134..102S}, and the solid lines indicate our selection criteria. Note that there were no known quasars fainter than $i=20.2$ mag here. In order to reduce the number of contaminants, we used slightly different criteria in three different magnitude ranges, $19.0<i<20.2$ mag, $20.2<i<20.8$ mag, and $20.8<i<21.3$ mag. All selection criteria are provided in Appendix \ref{App:sel}.

The selection of quasar candidates at $4.5<z<5.5$ ($riz$ candidates) was based on the $i-z$ versus $r-i$ diagram (the right column of Figure \ref{fig:selection}). We also used slightly different criteria in the three magnitude ranges and the criteria are shown in Appendix \ref{App:sel}. These criteria are very similar to those used in the literature \citep[e.g.,][]{2002AJ....123.2945R,2013ApJ...768..105M,2016ApJ...819...24W,2016ApJ...829...33Y}.

The target selection in Stripe 82 is straightforward. The combined images and photometric catalogs were available in the archive, and the data were much deeper \citep{2014ApJ...794..120A}. We selected candidates down to $i=21.5$ mag and included objects brighter than $i=19.0$ mag from the SDSS Query CasJobs online server (using `Run'$=106$ or 206 and `specObjID'$=0$). The overall selection criteria are very similar to Equations \ref{eq:overlap_gri_1} and \ref{eq:overlap_riz_1}. They are shown in Appendix \ref{App:sel}. The search area is $22^\mathrm{h}<\mathrm{R.A.}<4^\mathrm{h}$. We did not use the region of R.A. $<22^\mathrm{h}$ deg as the Galactic latitude becomes lower.

\subsection{SDSS-III Spectroscopic Observations}
Our targets were observed by the BOSS spectrograph in SDSS-III \citep{2011AJ....142...72E,2013AJ....145...10D}. The BOSS main survey covered $\sim$10,000 deg$^2$ in the north and south galactic caps and was completed in 2014 \citep{2015ApJS..219...12A}. The BOSS quasar survey mainly focused on quasars at $2.2<z<3.5$. Our program was selected as one of the ancillary programs to fill spare fibers. The SDSS bitmasks used in SDSS targeting can be found in the website\footnote{\url{https://www.sdss.org/dr12/algorithms/ancillary/boss/highz/}} of our program. From the selection procedure above, we obtained 4374 quasar candidates, including 3454 candidates from the overlap regions and 920 candidates from Stripe 82. A total of 3406 candidates were spectroscopically observed. The mean fraction of targets with spectroscopic observations reaches about 78\% and we will correct this incompleteness in Section \ref{sec:corr}.

\section{Results} \label{sec:results}

\subsection{Quasar Sample}

From the spectroscopic observations, we obtained 887 quasars. They have been included in the SDSS DR16 quasar catalogue \citep{2020ApJS..250....8L}. Their redshift distribution is shown in Figure \ref{fig:lowz}. This sample consists of 35 quasars at $z<3$ and 852 quasars at $z>3$. The $z>3$ sample (hereafter new sample) includes 652 quasars in the overlap regions and 200 quasars in Stripe 82. 

As we mentioned earlier, we did not observe the objects that had already been spectroscopically observed in SDSS I and II. Some of them also satisfy our target selection criteria. We recovered this quasar sample (hereafter the archival sample) as follows. For quasars in the overlap regions, we changed one criterion (using `specObjID'$!=0$) and repeated the selection procedure. For quasars in Stripe 82, we directly used the criteria to match quasars in the DR7 quasar catalogue \citep[hereafter DR7Q,][]{2010AJ....139.2360S}. We recovered a total of 346 quasars. Our final high-redshift sample is the combination of the archival and new samples, and consists of 1198 quasars at $z>3$ (Table \ref{table:total}).  

We measure the continuum properties of the quasars assuming a power-law shape $f_\lambda\propto\lambda^{\alpha_\lambda}$. We fit this power-law to the spectral regions with little line emission. The resultant slope $\alpha_\lambda$ distribution is shown in Figure \ref{fig:alpha}. The mean $\alpha_\lambda$ value is about $-1.1$, similar to previous measurements of high-redshift quasars \citep[e.g.,][]{2001AJ....121...31F,2001AJ....121.1232S}, but quite softer than the results from low-redshift works due to the short wavelength coverages for high-redshift quasars \citep{2001AJ....122..549V}. Continuum luminosity/magnitude $M_{1450}$ is also calculated in this step. Figure \ref{fig:M1450_z} shows the redshift and $M_{1450}$ distributions of the archival sample and new sample. The median value of $M_{1450}$ is around $-25.5$ mag. The new sample is about 1 magnitude deeper on average, so the QLF calculated in this paper will reach a lower luminosity compared to that from the SDSS single-epoch data. Table \ref{table:highz} lists our high-redshift quasar sample. 

\subsection{Area Coverage}

The area coverage of the SDSS overlap regions is complex. We use the Hierarchical Equal Area isoLatitude Pixelization \citep[HEALPix;][]{2005ApJ...622..759G} to estimate the effective area of the overlap regions. The basic idea is to pixelize the sky sphere into a mesh of quadrilateral pixels and the effective area is calculated by adding up all pixels that cover our data points.

For a given dataset, the starting resolution level of HEALPix is important for the area calculation. We follow \citet{2016ApJ...833..222J} and adopt HEALPix Level 10 {(i.e., 11.8 square arcminutes per pixel) as the best starting level for the overlap regions \citep[see Figure 5 in][]{2016ApJ...833..222J}. We classify all pixels into three categories. The first category consists of empty pixels that do not cover any objects, so they do not contribute to the effective coverage. The close neighbors to empty pixels are boundary pixels that will result in the uncertainty of the area calculation. The remaining pixels are all in the third category (hereafter non-boundary pixels). All non-boundary pixels at Level 10 contribute to the total effective area. We then gradually increase the resolution level for the boundary pixels. The resultant new pixels are once again classified into two categories, boundary and non-boundary pixels. The new non-boundary pixels are added to the total area and the new boundary pixels are refined again by increasing the resolution level. This procedure stops when the resolution roughly matches the average surface density of the data points. In this paper, we reach the best resolution at Level 12. 

\begin{figure}
\includegraphics[width=230 pt,height=200 pt]{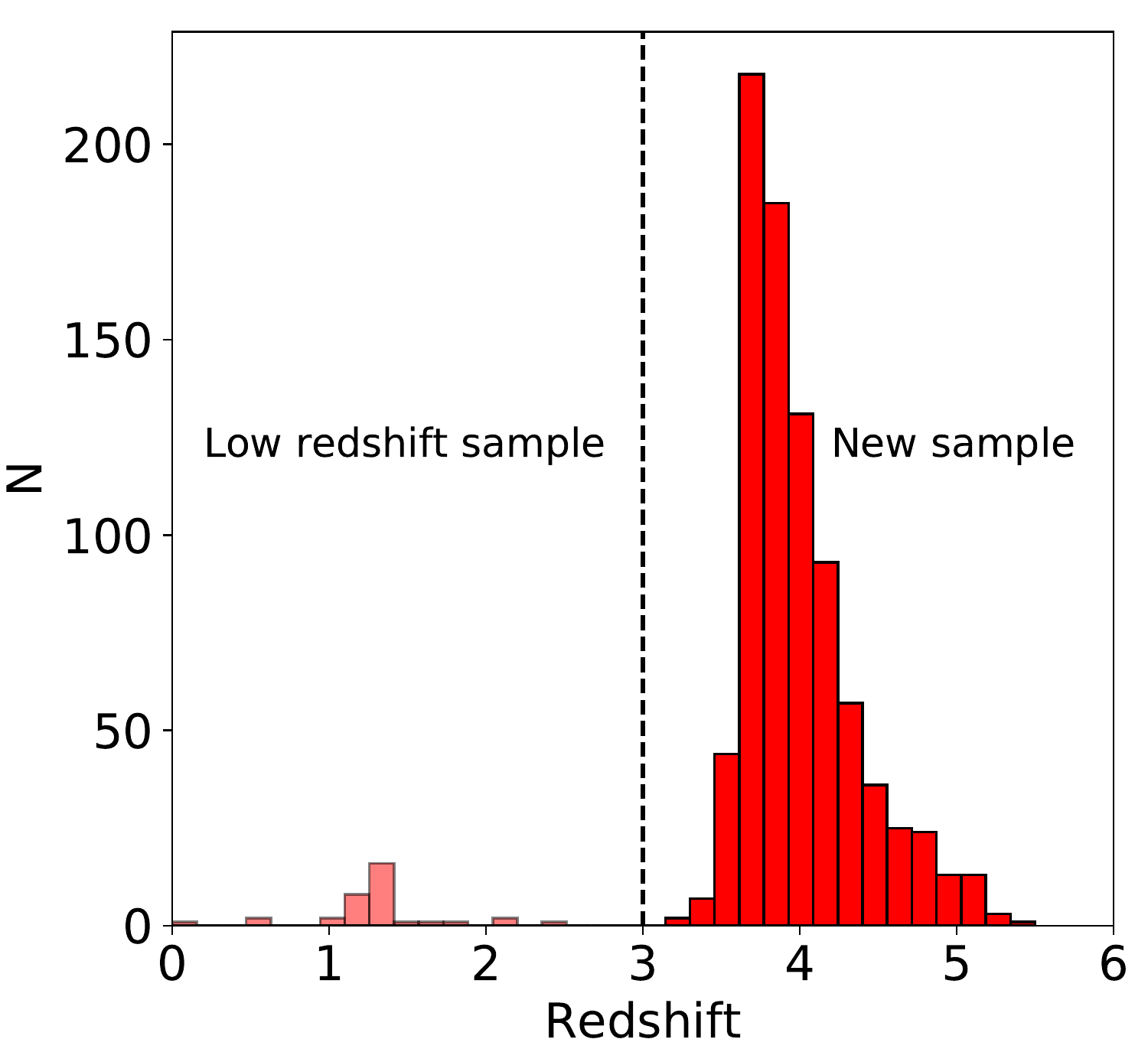}
\caption{Redshift distribution of the 887 quasars and 852 of them are at $z>3$ (the new sample). 
\label{fig:lowz}}
\end{figure}

From the above calculation, the total area of the overlap regions is $1292\pm266\ \rm{deg^2}$. The uncertainty of area coverage will be included in the measurement of the QLF. The calculation of the Stripe 82 area is very straightforward, since it is one rectangular piece of the sky. Its coverage is $225\ \rm{deg^2}$ with a negligible uncertainty. 

\begin{figure}
\epsscale{1.25}
\plotone{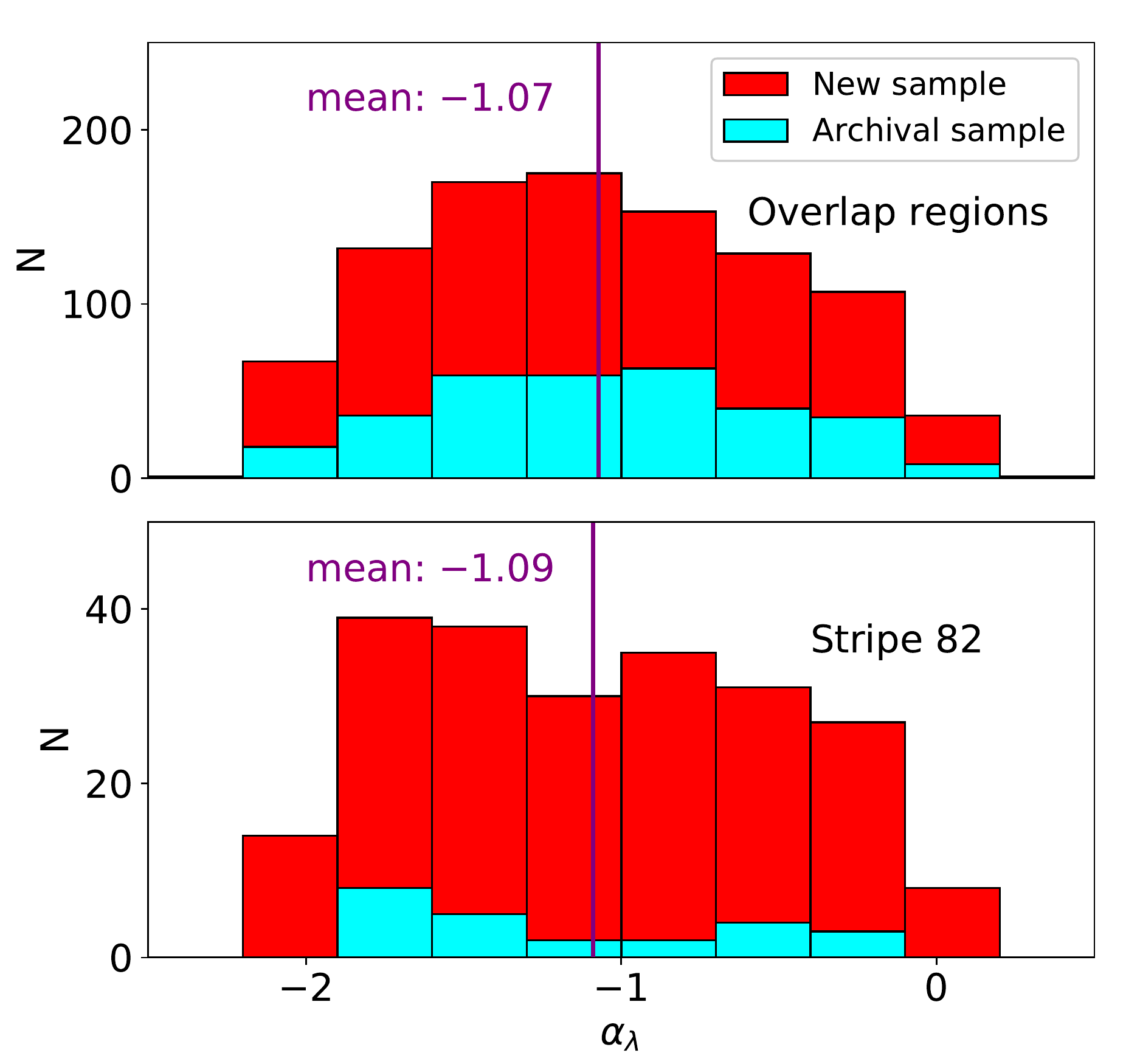}
\caption{Continuum slope $\alpha_{\lambda}$ distribution of the new (red) and  archival sample (cyan). The mean value of the continuum slope is about $-1.1$, which is similar to the previous results.
\label{fig:alpha}}
\end{figure}

\subsection{Sample Completeness} \label{sec:corr}

In this subsection we estimate our sample incompleteness that is critical to derive QLFs. The first incompleteness is from the fact that the BOSS survey did not observe all our targets. For example, about 82\% (71\%) of the targets in the $gri$ ($riz$) sample at $i<20.2$ mag were observed in the overlap regions. When we correct this incompleteness, we assume that the quasar fraction in the unobserved candidates is the same as that in the observed sample. The incompleteness slightly varies with the $i$-band magnitude. This variation is considered as the uncertainty of this incompleteness, and the results (1-3\%) are negligible.

\begin{figure*}
\epsscale{1.2}
\plotone{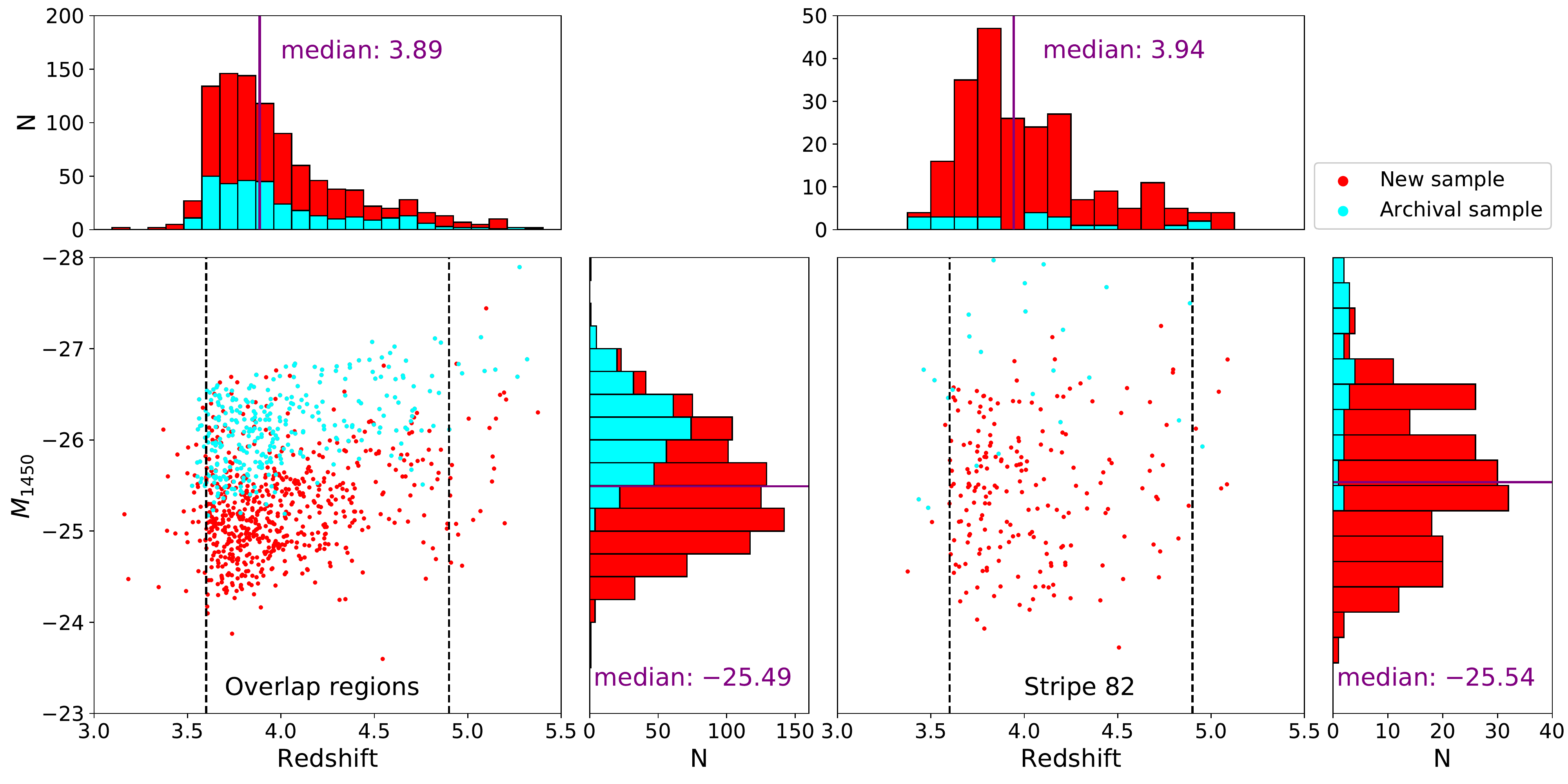}
\caption{Redshift and $M_{1450}$ distributions of the new sample (red) and the archival sample (cyan) at $3.0<z<5.5$. The median value of $M_{1450}$ is around $-25.5$ mag, which is about 1 to 2 mag deeper on average than the archival sample. The dots within the dashed lines ($3.6<z<4.9$) represent the sample used for our QLF measurements.
\label{fig:M1450_z}}
\end{figure*}

The second incompleteness arises from the morphological bias. The quasar candidates that we observed are point sources, but faint point sources with low signal-to-noise ratios can be misclassified as extended sources by the SDSS photometric pipeline. We correct this bias for the targets in the overlap regions (i.e., the single-epoch data). The Stripe 82 imaging data are much deeper and we assume that the targets in this region do not suffer from the morphological bias. We will see below that this assumption is reasonable. To estimate the bias for the single-epoch data, we use 27,593 point sources classified in Stripe 82. We divide the data into narrow magnitude bins. For each magnitude bin, we calculate the fraction of the objects that are misclassified as extended sources in the single-epoch data. The resultant fractions are from 0.04 at $19.0<i<19.6$ mag to 0.55 at $21.2<i<21.3$ mag. There is a clear relation between the fraction and brightness. The fraction in the brightest range is nearly zero, suggesting little bias for the targets in Stripe 82. These fractions are considered as sample incompletenesses in individual bins, and will be included when we calculate QLFs. In order to estimate the uncertainty of the incompleteness, we resample the data 1000 times. For each time, we randomly select 500 point sources per bin to estimate the incompleteness and finally regard the standard deviation as uncertainty. The resultant uncertainties (1-2\%) are negligible.

The next incompleteness comes from our color selection criteria, i.e., the color cuts introduced in Section \ref{sec:selection}. This incompleteness is described by a selection function, the probability that a quasar with a given magnitude ($M_{1450}$), redshift ($z$), and intrinsic spectral energy distribution (SED) meets the color selection criteria. We calculate the average selection probability $p_s(M_{1450},z)$ by assuming that intrinsic SEDs have certain distributions. Following the procedure in \citet{1999AJ....117.2528F} and \citet{2013ApJ...768..105M}, we use a simulation tool to estimate the selection function $p_s(M_{1450},z)$. \citet{2013ApJ...768..105M} updated the quasar spectral model in the code based on the colors of $\sim$ 60,000 quasars at $2.2 < z < 3.5$ from SDSS III \citep{2012ApJS..199....3R}. This model consists of a broken power-law continuum, most emission lines, an IGM absorption model, and an Fe emission template. It also accounts for the Baldwin effect. \citet{2016ApJ...833..222J} and \citet{2016ApJ...829...33Y} extended this model to higher redshifts with the assumption that quasar SEDs do not evolve with redshift. 

Based on the model of \citet{2016ApJ...829...33Y}, we generate a sample of simulated quasars with proper photometric errors. We follow the procedure in \citet{2016ApJ...833..222J} to add photometric errors. We withdraw a large representative sample of point sources in the overlap regions and Stripe 82, and derive error distributions as a function of magnitude in the $u,g,r,i,z$ bands. Finally, we construct a grid of 2,000,000 mock quasars in a redshift range of $3.5<z<5.5$ and a luminosity range of $-27.5<M_{1450}<-23.5$, with step sizes of $\Delta M=0.02$ and $\Delta z=0.02$. Then we calculate the selection function $p_s (M_{1450},z)$, the fraction of simulated quasars that meet our selection criteria. Figure \ref{fig:self_overlap} shows the selection functions for the overlap regions and Stripe 82. We estimate the uncertainty of the selection function using the same method as we did for the morphological incompleteness, and the result is around 1\%. As we will see that the uncertainties of the incompleteness corrections are negligibly small compared to other uncertainties, so they are not included in the following QLF calculations.

\begin{deluxetable*}{cccccccc}  
\tablecaption{High-Redshift Quasar Catalogue} 
\tablewidth{0pt}
\tablehead{ Quasar(SDSS)  & Redshift & $g$            & $r$            & $i$            & $z$            & $M_{1450}$ & Regions\tablenotemark{\scriptsize$1$}}
\startdata    \label{table:highz}
J102859.78$+$434656.4       & 3.16     & 21.30$\pm$0.04 & 20.42$\pm$0.02 & 20.37$\pm$0.03 & 20.26$\pm$0.10 & $-25.18$   & Overlap \\
J102513.31$+$350325.0       & 3.18     & 22.03$\pm$0.06 & 20.96$\pm$0.03 & 20.98$\pm$0.05 & 20.89$\pm$0.13 & $-24.47$   & Overlap\\
J152126.66$+$191816.9       & 3.35     & 21.70$\pm$0.04 & 21.02$\pm$0.03 & 21.24$\pm$0.05 & 21.21$\pm$0.18 & $-24.39$   & Overlap \\
J102940.93$+$100410.9       & 3.37     & 20.13$\pm$0.02 & 19.51$\pm$0.02 & 19.44$\pm$0.02 & 19.35$\pm$0.05 & $-26.11$   & Overlap \\
J113957.54$+$011458.5       & 3.39     & 21.37$\pm$0.04 & 20.73$\pm$0.03 & 20.62$\pm$0.04 & 20.36$\pm$0.12 & $-25.00$   & Overlap \\
... \\
J021149.15$-$010956.7       & 3.38     & 21.91$\pm$0.03 & 21.09$\pm$0.01 & 21.03$\pm$0.02 & 20.97$\pm$0.05 & $-24.56$   & S82 \\
J235219.08$-$000012.1       & 3.44     & 21.10$\pm$0.04 & 20.24$\pm$0.05 & 20.20$\pm$0.03 & 20.30$\pm$0.14 & $-25.35$   & S82 \\
J223843.56$+$001647.9       & 3.46     & 19.78$\pm$0.02 & 18.79$\pm$0.02 & 18.71$\pm$0.02 & 18.48$\pm$0.04 & $-26.77$   & S82 \\
J234548.18$+$000548.4       & 3.49     & 21.12$\pm$0.05 & 20.29$\pm$0.05 & 20.27$\pm$0.04 & 20.19$\pm$0.18 & $-25.26$   & S82 \\
J233101.65$-$010604.2       & 3.51     & 22.31$\pm$0.05 & 20.66$\pm$0.01 & 20.25$\pm$0.01 & 20.15$\pm$0.03 & $-25.10$   & S82 \\
... \\
\enddata
\tablecomments{This table is available in its entirety in the machine-readable format. The magnitudes for the overlap regions are combined magnitudes as introduced before. The magnitudes for the Stripe 82 are from the SDSS Query CasJobs online server or DR7Q depending on whether thay have previous spectroscopic observations. All the magnitudes are expressed in the AB system and have been corrected for the extinctions.}
\tablenotetext{$\scriptsize1$}{`Overlap': the overlap regions, `S82': Stripe 82.}
\end{deluxetable*} 

\subsection{Binned QLFs}

We use a traditional $1/V_\mathrm{a}$ method \citep{1980ApJ...235..694A} to derive the binned differential QLFs. The available volume for a quasar with absolute magnitude $M$ and redshift $z$ in a magnitude bin $\Delta M$ and a redshift bin $\Delta z$ is 
\begin{equation} \label{eq:va}
V_\mathrm{a}=\iint_{\Delta M\Delta z}^{}p(M,z)\frac{dV}{dz}dzdM,
\end{equation}
where $p(M,z)$ is the final selection function that includes all incompleteness corrections discussed above. 

In general, the binned QLF and its statistical uncertainty can be expressed as 
\begin{equation} \label{eq:binned_qlf}
\Phi(M,z)=\sum\frac{1}{V_\mathrm{a}^i}, \sigma(\Phi)=[\sum(\frac{1}{V_\mathrm{a}^i})^2]^{1/2},
\end{equation}
where the sum is over all quasars in each bin. When a density approaches the Poisson limit, its uncertainty is corrected using  Equation 7 in \citet{1986ApJ...303..336G}. 

We divide our sample into several luminosity bins and redshift bins, and we focus on three redshift ranges, $3.6<z<4.0, 4.0<z<4.5$, and $4.5<z<4.9$. The magnitude limits in each redshift range are determined by the faintest and/or brightest quasars and the selection functions. The binned QLF results are listed in Table \ref{table:qlf} and also displayed in Figure \ref{fig:binned_qlf} as the blue circles (Overlap regions) and red circles (Stripe 82). The horizontal locations of the symbols are at the centers of each magnitude bin, and the horizontal bars indicate the magnitude coverage ranges. The binned QLFs calculated for the two datasets are consistent within $1\sigma$.

\subsection{Maximum Likelihood Fitting}

We combine the two datasets from the overlap regions and Stripe 82 and derive a parametric QLF using the maximum likelihood method \citep{1983ApJ...269...35M}. This method aims to minimize the function $S$, which is equal to $-2\ln L$, where L is the likelihood function:
\begin{equation} \label{eq:ML}
\begin{split}
S=-&2\sum \ln[\Phi(M_i,z_i)p(M_i,z_i)]  \\
  +&2\int_{\Delta M}\int_{\Delta z}\Phi(M,z)p(M,z)\frac{dV}{dz}dzdM,
\end{split}
\end{equation}
where $p(M,z)$ includes all the incompleteness corrections discussed above. The first term is the sum over all observed quasars in the sample. The second term is integrated over the whole magnitude and redshift range of the sample. It represents the total number of expected quasars for a given luminosity function. The confidence intervals are determined from the logarithmic likelihood function using a $\chi^2$ distribution of $\Delta S$ ($\mathbf{=S-S_{min}}$) \citep{1976ApJ...208..177L}. 

We choose a double power-law form \citep{2000MNRAS.317.1014B} as the parametric QLF model:
\begin{equation} \label{eq:double_power}
\Phi_{\mathrm{par}(M,z)}=\frac{\Phi^*}{10^{0.4(\alpha+1)(M-M^*)}+10^{0.4(\beta+1)(M-M^*)}},
\end{equation}
where $\alpha$ and $\beta$ are the faint- and bright-end slopes, $M^*$ is the characteristic magnitude (or break magnitude), and $\Phi^*$ is the density normalization. We assume that these parameters do not change in small redshift ranges, such as the ranges considered here. We will discuss the QLF evolution in Section \ref{sec:evole}. We perform a grid search to determine the best-fit results and the confidence intervals. The grid resolutions of log$\Phi^*$, $M^*$, and $\alpha$ are 0.05, 0.05, and 0.1, respectively. There is a strong degeneracy between $M^*$ and $\alpha$, so we set a bright limit of $-28.0$ mag for $M^*$. The best-fit results are listed in Table \ref{table:fit}. 

Figure \ref{fig:binned_qlf} shows the results in three redshift ranges. The open circles denote the data points (some of the faintest bins) that have very low completeness and significantly deviate from the general trend. It is unclear what causes this deviation. This has been frequently seen in previous studies and is likely due to some unknown selection effects. We did not use these data points in the above calculation. Our sample covers a limited range of luminosity, so it is not able to constrain both slopes $\alpha$ and $\beta$. Therefore, we use three fixed values for $\beta$ in each redshift range (see Figure \ref{fig:binned_qlf}) and derive the other three parameters. The best-fitted $\alpha$ values are about $-1.8$ at $3.6<z<4.9$, indicating that the results for different $\beta$ values and different redshift ranges are not significantly different. In addition, most of the best-fitted $M^*$ values for three redshift ranges are lower than $-27$ (see Table \ref{table:fit}), making the double power-law model degenerating into a single power-law model. These results suggest that our sample alone is not enough to constrain all parameters in the above QLF model. In the next section, we will combine our binned QLFs with some results in the literature.

\begin{figure}
\epsscale{1.2}
\plotone{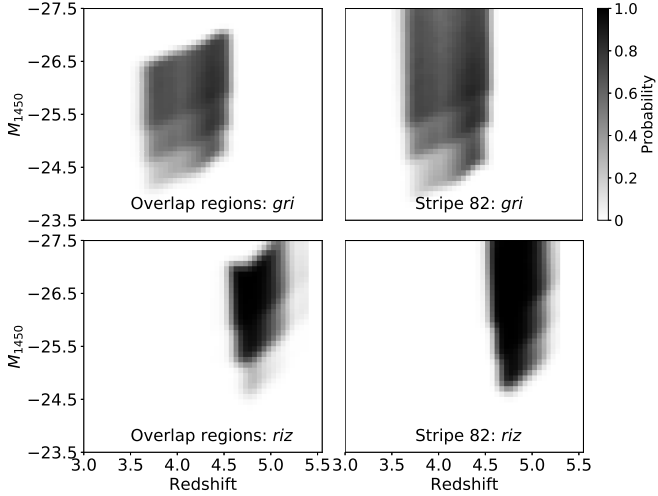}
\caption{Quasar selection functions $p_s(M_{1450},z)$ for the $gri$ criteria (upper panels) and $riz$ criteria (lower panels). Left: selection functions for the overlap regions. Right: selection functions for Stripe 82. 
\label{fig:self_overlap}}
\end{figure} 

\begin{deluxetable}{ccccccc}  
\tablecaption{Binned QLF}
\tablewidth{0pt}
\tablehead{$M_{1450}$ & $N$  & log$\Phi$\tablenotemark{\scriptsize$1$} & $\Delta\Phi$\tablenotemark{\scriptsize$2$} & Regions} 
\startdata    \label{table:qlf}
$3.6<z<4.0$ \\
\tableline
$-24.30$              & 49                 & $-6.62$                    & 62.92                         & Overlap \\   
$-24.80$              & 108                & $-6.56$                    & 64.38                         & Overlap \\ 
$-25.15$              & 107                & $-6.64$                    & 53.68                         & Overlap \\ 
$-25.45$              & 97                 & $-6.87$                    & 31.87                         & Overlap \\ 
$-25.80$              & 93                 & $-7.13$                    & 17.54                         & Overlap \\
$-26.15$              & 64                 & $-7.19$                    & 16.01                         & Overlap \\ 
$-26.70$              & 55                 & $-7.36$                    & 11.31                         & Overlap \\ 
$-24.20$              & 8                  & $-7.01$                    & 47.99                         & S82 \\   
$-24.65$              & 12                 & $-6.72$                    & 73.01                         & S82 \\ 
$-25.00$              & 18                 & $-6.91$                    & 36.44                         & S82 \\ 
$-25.45$              & 27                 & $-6.95$                    & 26.06                         & S82 \\   
$-26.05$              & 35                 & $-7.03$                    & 18.57                         & S82 \\ 
$-26.80$              & 17                 & $-7.40$                    & 12.35                         & S82 \\ 
\tableline
$4.0<z<4.5$ \\
\tableline
$-24.40$              & 9                  & $-7.23$                    & 29.92                         & Overlap \\  
$-24.85$              & 52                 & $-7.08$                    & 21.6                         & Overlap \\  
$-25.25$              & 40                 & $-7.22$                    & 16.73                         & Overlap \\  
$-25.55$              & 42                 & $-7.29$                    & 13.95                         & Overlap \\  
$-25.85$              & 33                 & $-7.53$                    & 8.62                          & Overlap \\  
$-26.25$              & 44                 & $-7.72$                    & 5.15                          & Overlap \\  
$-26.85$              & 21                 & $-8.00$                    & 3.43                          & Overlap \\  
$-24.30$              & 9                  & $-7.01$                    & 44.45                         & S82 \\
$-24.75$              & 14                 & $-7.24$                    & 19.71                         & S82 \\
$-25.30$              & 14                 & $-7.44$                    & 12.47                         & S82 \\
$-25.80$              & 9                  & $-7.52$                    & 13.91                         & S82 \\
$-26.25$              & 9                  & $-7.63$                    & 10.73                         & S82 \\
$-27.00$              & 9                  & $-7.93$                    & 5.42                          & S82 \\
\tableline
$4.5<z<4.9$ \\
\tableline
$-24.90$              & 13                 & $-7.54$                    & 12.04                         & Overlap \\  
$-25.65$              & 27                 & $-7.61$                    & 7.78                          & Overlap \\  
$-26.15$              & 21                 & $-7.93$                    & 4.07                          & Overlap \\  
$-26.60$              & 12                 & $-8.18$                    & 2.85                          & Overlap \\ 
$-27.15$              & 8                  & $-8.30$                    & 2.73                          & Overlap \\ 
$-24.85$              & 7                  & $-7.56$                    & 14.81                         & S82 \\  
$-25.60$              & 6                  & $-7.76$                    & 10.45                         & S82 \\  
$-26.30$              & 5                  & $-8.06$                    & 5.97                          & S82 \\  
$-27.10$              & 4                  & $-8.17$                    & 5.36                          & S82 \\  
\tableline
\enddata
\tablenotetext{$\scriptsize1$}{$\Phi$ is in units of $\mathrm{Mpc}^{-3}\ \mathrm{mag}^{-1}$.}
\tablenotetext{$\scriptsize2$}{$\Delta\Phi$ is in units of $10^{-9}\ \mathrm{Mpc}^{-3}\ \mathrm{mag}^{-1}$.}
\end{deluxetable}

\section{Discussion}   \label{sec:dis}
\subsection{Comparison with Previous Work}
In Figure \ref{fig:binned_qlf_compare} we show a collection of previous QLF measurements at $3.6<z<4.9$ \citep{Richards_2006,2013ApJ...768..105M,2016ApJ...829...33Y,Boutsia_2018,McGreer_2018,Schindler_2019,2019MNRAS.488.1035K,Kim_2020,2021arXiv210310446B}. All data points have been scaled to $z=3.8, 4.25$, or $4.7$ using the density evolution model of \citet{Schindler_2019} with $\gamma=-0.38$ (e.g., $\mathbf{\Phi(z=3.8)=\Phi(z)*10^{-\gamma(z-3.8)}}$). \citet{Richards_2006} constructed a sample of 15,343 quasars at $0<z<5$ in 1,622 $\mathrm{deg}^2$ from SDSS DR3, with a small fraction of quasars at $z>3.6$. Our sample is roughly 1.5 mag deeper than this sample. \citet{Boutsia_2018} focused on faint quasars at $3.6<z<4.2$ in the COSMOS field. \citet{Schindler_2019} built a sample of very luminous quasars at $2.8<z<4.5$ to constrain the bright-end slope $\beta$ of the QLF.  \citet{2021arXiv210310446B} identified 58 bright quasars at $3.6<z<4.2$ and the brightest ones reach $M_{1450}=-29.5$ mag. \citet{2019MNRAS.488.1035K} combined multiple datasets from previous surveys and generated a sample of more than 80,000 quasars, with a small fraction of quasars at $z>3.6$. They used this sample to derive QLFs and study quasar evolution from $z=7.5$ to 0. In the bottom panel of Figure \ref{fig:binned_qlf_compare}, we particularly compared our results with previous QLF measurements at $z\sim5$ \citep[e.g.,][]{2013ApJ...768..105M,2016ApJ...829...33Y,McGreer_2018,Kim_2020}. We did not include samples with no or very few spectroscopic observations \citep[e.g.,][]{2018PASJ...70S..34A,2020ApJ...904...89N}.

\begin{figure}
\epsscale{1.0}
\plotone{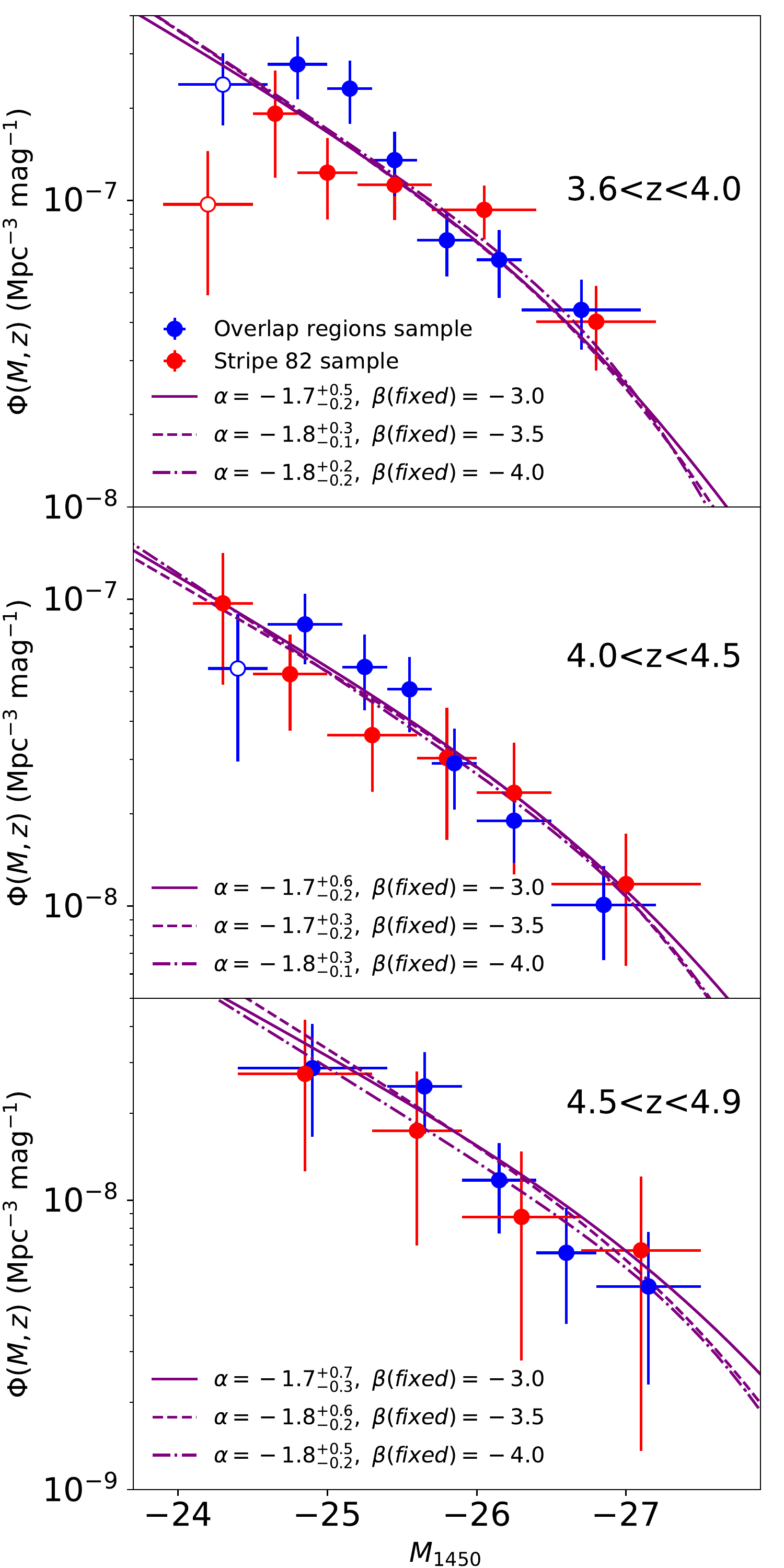} 
\caption{The binned QLFs at $3.6<z<4.9$ derived from our sample. The open circles denote the data points with very low completeness that are not used for our parametric QLF fitting. The purple lines show the best-fitted QLFs.
\label{fig:binned_qlf}}
\end{figure}

\begin{figure}
\epsscale{1.2}
\plotone{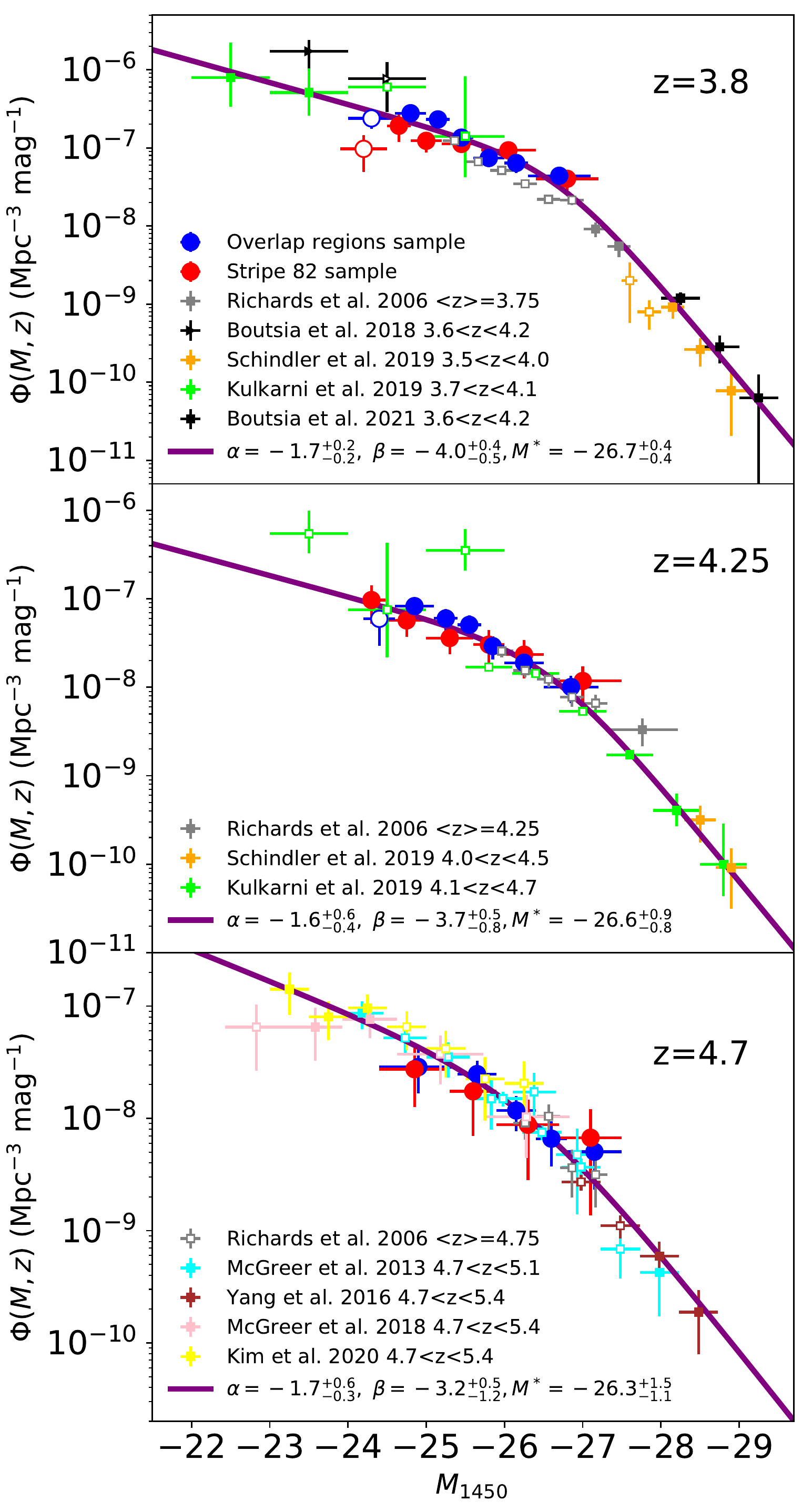}
\caption{QLFs at $3.6<z<4.9$ from the combination of our work and the literature results. All data points from the literature have been scaled to $z=3.8$, 4.25, and 4.7, respectively, by adopting the density evolution model of \citet{Schindler_2019} with $\gamma=-0.38$. The solid purple lines are the best-fit QLFs. The open symbols are not used in our fitting process (see details in the text).}     \label{fig:binned_qlf_compare}
\end{figure}

\begin{deluxetable*}{ccccccccccc}  
\tablecaption{Parameters of the Best-Fits} 
\tablewidth{0pt}             
\tablehead{   & Sample\tablenotemark{\scriptsize$1$} & $\mathrm{log}\Phi^*_0$  & $M^*_0$                  & $\alpha_0$              & $\beta_0$               & $k_{\Phi}$               & $k_{M}$                 & $k_{\alpha}$            & $k_{\beta}$            & ${\chi^2}_{\nu}$} 
\startdata    \label{table:fit}
$3.6<z<4.0$ \\
\tableline
Fixed $\beta$ & O+S                                  & $-7.5^{+0.4}_{-0.5}$ & $-27.3^{+0.7}_{-0.8}$ & $-1.8^{+0.2}_{-0.2}$    & $-4.0$                  & ...                      & ...                     & ...                     & ...                    & ... \\   
Fixed $\beta$ & O+S                                  & $-7.5^{+0.5}_{-0.4}$ & $-27.3^{+1.0}_{-0.7}$ & $-1.8^{+0.3}_{-0.1}$    & $-3.5$                  & ...                      & ...                     & ...                     & ...                    & ... \\   
Fixed $\beta$ & O+S                                  & $-7.3^{+0.5}_{-0.5}$ & $-27.0^{+1.1}_{-1.0}$ & $-1.7^{+0.5}_{-0.2}$    & $-3.0$                  & ...                      & ...                     & ...                     & ...                    & ... \\   
Best fit      & O+S+L                                & $-7.2^{+0.3}_{-0.3}$ & $-26.7^{+0.4}_{-0.4}$ & $-1.7^{+0.2}_{-0.2}$ & $-4.0^{+0.4}_{-0.5}$ & ...                      & ...                     & ...                     & ...                    & ... \\  
\tableline
$4.0<z<4.5$ \\
\tableline 
Fixed $\beta$ & O+S                                  & $-8.1^{+0.5}_{-0.2}$ & $-27.7^{+1.0}_{-0.3}$ & $-1.8^{+0.3}_{-0.1}$    & $-4.0$                  & ...                      & ...                     & ...                     & ...                    & ... \\   
Fixed $\beta$ & O+S                                  & $-7.9^{+0.5}_{-0.4}$ & $-27.4^{+0.9}_{-0.7}$ & $-1.7^{+0.3}_{-0.2}$    & $-3.5$                  & ...                      & ...                     & ...                     & ...                    & ... \\   
Fixed $\beta$ & O+S                                  & $-7.9^{+0.7}_{-0.4}$ & $-27.5^{+1.6}_{-0.6}$ & $-1.7^{+0.6}_{-0.2}$    & $-3.0$                  & ...                      & ...                     & ...                     & ...                    & ... \\   
Best fit      & O+S+L                                & $-7.6^{+0.5}_{-0.5}$ & $-26.6^{+0.9}_{-0.8}$ & $-1.6^{+0.6}_{-0.4}$    & $-3.7^{+0.5}_{-0.8}$ & ...                      & ...                     & ...                     & ...                    & ... \\  
\tableline
$4.5<z<4.9$ \\
\tableline 
Fixed $\beta$ & O+S                                  & $-8.5^{+0.7}_{-0.1}$ & $-28.0^{+1.4}_{-0.0}$ & $-1.8^{+0.5}_{-0.2}$    & $-4.0$                  & ...                      & ...                     & ...                     & ...                    & ... \\  
Fixed $\beta$ & O+S                                  & $-8.4^{+0.8}_{-0.1}$ & $-27.9^{+1.5}_{-0.1}$ & $-1.8^{+0.6}_{-0.2}$    & $-3.5$                  & ...                      & ...                     & ...                     & ...                    & ... \\  
Fixed $\beta$ & O+S                                  & $-8.3^{+0.8}_{-0.2}$ & $-27.9^{+1.8}_{-0.1}$ & $-1.7^{+0.7}_{-0.3}$    & $-3.0$    & ...                      & ...                     & ...                     & ...                    & ... \\    
Best fit      & O+S+L                                & $-7.7^{+0.7}_{-0.7}$ & $-26.3^{+1.5}_{-1.1}$ & $-1.7^{+0.6}_{-0.3}$ & $-3.2^{+0.5}_{-1.2}$ & ...                      & ...                     & ...                     & ...                    & ... \\   
\tableline
$3.6<z<4.9$ \\
Case 1        & O+S+L                   			 & $-7.4^{+0.2}_{-0.2}$ & $-27.0^{+0.4}_{-0.3}$ & $-1.9^{+0.1}_{-0.1}$ & $-4.1^{+0.4}_{-0.5}$ & $-0.7^{+0.1}_{-0.1}$  & ...                     & ...                     & ...                    & 0.65  \\ 
Case 2        & O+S+L                   			 & $-7.4^{+0.2}_{-0.2}$ & $-27.1^{+0.4}_{-0.3}$ & $-2.0^{+0.1}_{-0.1}$ & $-4.5^{+0.5}_{-0.6}$ & $-0.7^{+0.1}_{-0.1}$  & ...                     & ...                     & $0.7^{+0.5}_{-0.5}$ & 0.59  \\
Case 3        & O+S+L                 				 & $-7.3^{+0.2}_{-0.2}$ & $-26.9^{+0.4}_{-0.3}$ & $-2.0^{+0.1}_{-0.1}$ & $-4.2^{+0.4}_{-0.5}$ & $-0.9^{+0.2}_{-0.2}$  & $-0.4^{+0.3}_{-0.4}$ & ...                     & ...                    & 0.61 \\
Case 4        & O+S+L                 			     & $-6.8^{+0.3}_{-0.3}$ & $-26.3^{+0.5}_{-0.5}$ & $-1.5^{+0.3}_{-0.2}$ & $-4.0^{+0.5}_{-0.6}$ & $-1.5^{+0.4}_{-0.3}$  & $-1.1^{+0.6}_{-0.4}$ & $-0.5^{+0.2}_{-0.2}$ & $0.1^{+0.8}_{-0.9}$ & 0.48 \\
\enddata
\tablenotetext{$\scriptsize1$}{`O+S': We combine the quasar samples from the overlap regions and Stripe 82; `O+S+L': We combine the observed binned QLFs from this work and the literature.}
\end{deluxetable*}

Figure \ref{fig:binned_qlf_compare} shows that our results are generally consistent with previous measurements. In the top panel, our luminosity coverage for $z=3.8$ partly overlap with the luminosities covered by \citet{Richards_2006}, \citet{Boutsia_2018}, and \citet{2019MNRAS.488.1035K}. In this overlap range, the binned QLFs from different studies roughly agree with each other. Our binned QLFs at $-26.5<M_{1450}<-25.5$ are about 1.5 times the results of \citet{Richards_2006}. It is unclear whether their results were underestimated or our results were overestimated. In the middle panel for $z=4.25$, our result is well consistent with the previous results from \citet{Richards_2006} and \citet{2019MNRAS.488.1035K} except two high data points from \citet{2019MNRAS.488.1035K}. The bottom panel shows several studies of the QLF at $z=4.7$ and most of these results are consistent with ours within 1 $\sigma$ level. It is worth noting that discrepancies are relatively larger at the faint- and bright- ends where uncertainties are also significantly large.

\subsection{Quasar Evolution at High Redshift} \label{sec:evole}

\begin{figure}
\epsscale{1.1}
\plotone{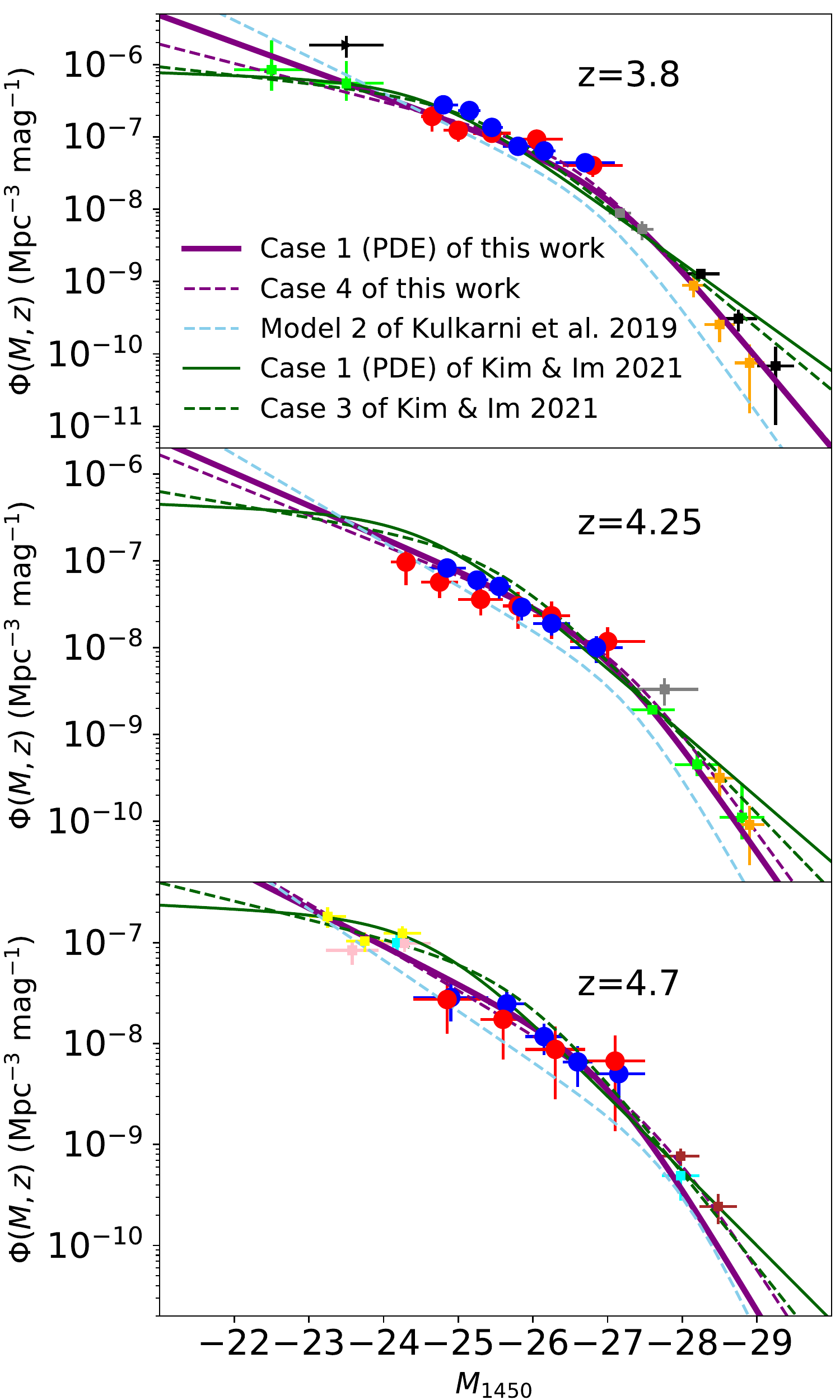}
\caption{QLF evolution at $z\sim3-5$. All the data points have been scaled to $z=3.8, 4.25, 4.7$ respectively by adopting the density evolution model in this work with $k_{\Phi}=-0.6$. The purple lines denote the best-fit QLFs in this work. The empirical models of \citet{2019MNRAS.488.1035K} and \citet{2021arXiv210306265K} are shown as the dark green and sky blue lines, respectively. The solid lines denote PDE models and the dashed lines denote other models.}     \label{fig:qlf_evol}
\end{figure}

In order to better constrain the shape of QLFs, we combine our QLF measurements with the results from some previous studies \citep{Richards_2006,2013ApJ...768..105M,2016ApJ...829...33Y,Boutsia_2018,McGreer_2018,Schindler_2019,2019MNRAS.488.1035K,Kim_2020,2021arXiv210310446B}. We caution that these previous samples are not fully independent, because the same quasars were used in some samples (thought analyzed with different methods). For the previous results, detailed selection functions for individual samples are not publicly available, so we directly use their binned QLFs. The derived parameters from binned data may be subject to some biases \citep[e.g.,][]{1997AJ....113.1517L,2000MNRAS.311..433P}. To reduce the impact from potential biases, we only use our results in the luminosity ranges that our sample covers. In addition, we exclude those points with very low completeness. The points that are not used in the fitting are shown as the open symbols in Figure \ref{fig:binned_qlf_compare} and the purple line in each panel denotes the best-fit QLF. We fit the observed QLF data points ($\Phi_{\mathrm{obs}}$) using the maximum likelihood estimation. We use the logarithmic likelihood function $\mathcal{L}$:
\begin{equation} \label{eq:ML_MCMC}
\mathcal{L}=-\frac{1}{2}\sum[(\Phi_{\mathrm{obs}}-\Phi_{\mathrm{par}})^2/\sigma^2+\mathrm{ln}(2\pi\sigma^2)],
\end{equation}
where $\sigma=\sqrt{\sigma_{\mathrm{obs}}^2+\sigma_{\mathrm{extra}}^2}$, $\sigma_{\mathrm{obs}}$ is the 1$\sigma$ uncertainty of $\Phi_{\mathrm{obs}}$ from the literature, and $\sigma_{\mathrm{extra}}$ is calculated from the magnitude bin size by the error propagation. The details of $\sigma_{\mathrm{extra}}$ and the chosen priors on the parameters are shown in Appendix \ref{App:MLE}. We use the emcee Python package \footnote{\url{https://emcee.readthedocs.io/en/stable/}} \citep{2013PASP..125..306F} for the Markov chain Carlo sampling of the QLF parameters. The best-fit results and uncertainties are estimated based on the 16th, 50th and 84th percentiles of the samples in the marginalized distributions. Figure \ref{fig:binned_qlf_compare} shows the best-fit QLFs and parameters are listed in Table \ref{table:fit}. 

The best-fit values at $z=3.8$ are $\alpha=-1.7^{+0.2}_{-0.2}$, $\beta=-4.0^{+0.4}_{-0.5}$, and $M^*=-26.7^{+0.4}_{-0.4}$. The $\beta$ and $M^*$ values are consistent with the results of \citet{Boutsia_2018} ($\beta=-4.025^{+0.575}_{-0.425}$ and $M^*=-26.50^{+0.85}_{-0.60}$) with reduced errors. The binned QLF data points of \citet{Boutsia_2018} are higher than previous measurements, resulting in a steeper faint-end slope $\alpha$ and a higher density $\Phi^*$ than our results. The QLF at $z=4.25$ is also well constrained by our results. As mentioned earlier, the faintest binned QLF calculated by \citet{2019MNRAS.488.1035K} is very high, so they obtained a steep faint-end slope $\alpha=-2.20^{+0.16}_{-0.14}$. Our slope $\alpha$ is slightly flatter. 

For the QLF at $z=4.7$, we adopt the results with $\alpha=-1.7^{+0.6}_{-0.3}$ and $\beta=-3.2^{+0.5}_{-1.2}$ as our best fit. We combined the data from \citet{Kim_2020} and \citet{McGreer_2018} for the faint end and the data from \citet{2013ApJ...768..105M} and \citet{2016ApJ...829...33Y} for the bright end. In this high-redshift range, the current sample size is still small and thus the above measurements are associated with relatively large uncertainties.

Based on the measurements from the above subsection, we explore quasar evolution at $3.5<z<5$. We use a linear function to describe the evolution of the four QLF parameters in this small redshift range:
\begin{equation} \label{eq:linear}
X(z)=X_0+k_{X}(z-3.5),
\end{equation}
where $X\in\{\mathrm{log}\Phi^*,\ M^*,\ \alpha,\ \beta\}$. Here we consider four cases:
\begin{enumerate}
\item[$\circ$] Case 1: A PDE model where only $\Phi^*$ evolves, i.e., we fit 5 parameters: $\mathrm{log}\Phi^*_0,\ M^*_0,\ \alpha_0,\ \beta_0,\ k_{\Phi}$.
\item[$\circ$] Case 2: We allow $\Phi^*$ and $\beta$ to evolve, i.e., we fit 6 parameters: $\mathrm{log}\Phi^*_0,\ M^*_0,\ \alpha_0,\ \beta_0,\ k_{\Phi},\ k_{\beta}$.
\item[$\circ$] Case 3: A LEDE model. We allow $\Phi^*$ and $M^*$ to evolve, i.e., we fit 6 parameters: $\mathrm{log}\Phi^*_0,\ M^*_0,\ \alpha_0,\ \beta_0,\ k_{\Phi},\ k_M$.
\item[$\circ$] Case 4: We allow all parameters to evolve, i.e., we fit 8 parameters: $\mathrm{log}\Phi^*_0,\ M^*_0,\ \alpha_0,\ \beta_0,\ k_{\Phi},\ k_M,\ k_{\alpha},\ k_{\beta}$.
\end{enumerate}

We fit these 4 models to the observed QLF data points using the Maximum likelihood estimation introduced earlier. We calculate the reduced $\chi^2$ as below,
\begin{equation} \label{eq:chinu}
\chi_\nu^2=\frac{\sum[(\Phi_{obs}-\Phi_{par})/\sigma]^2}{n-\nu},
\end{equation}
where $n$ is the number of data points and $\nu$ is the number of the free parameters. The results are listed in Table \ref{table:fit}.

\begin{figure}
\epsscale{1.19}
\plotone{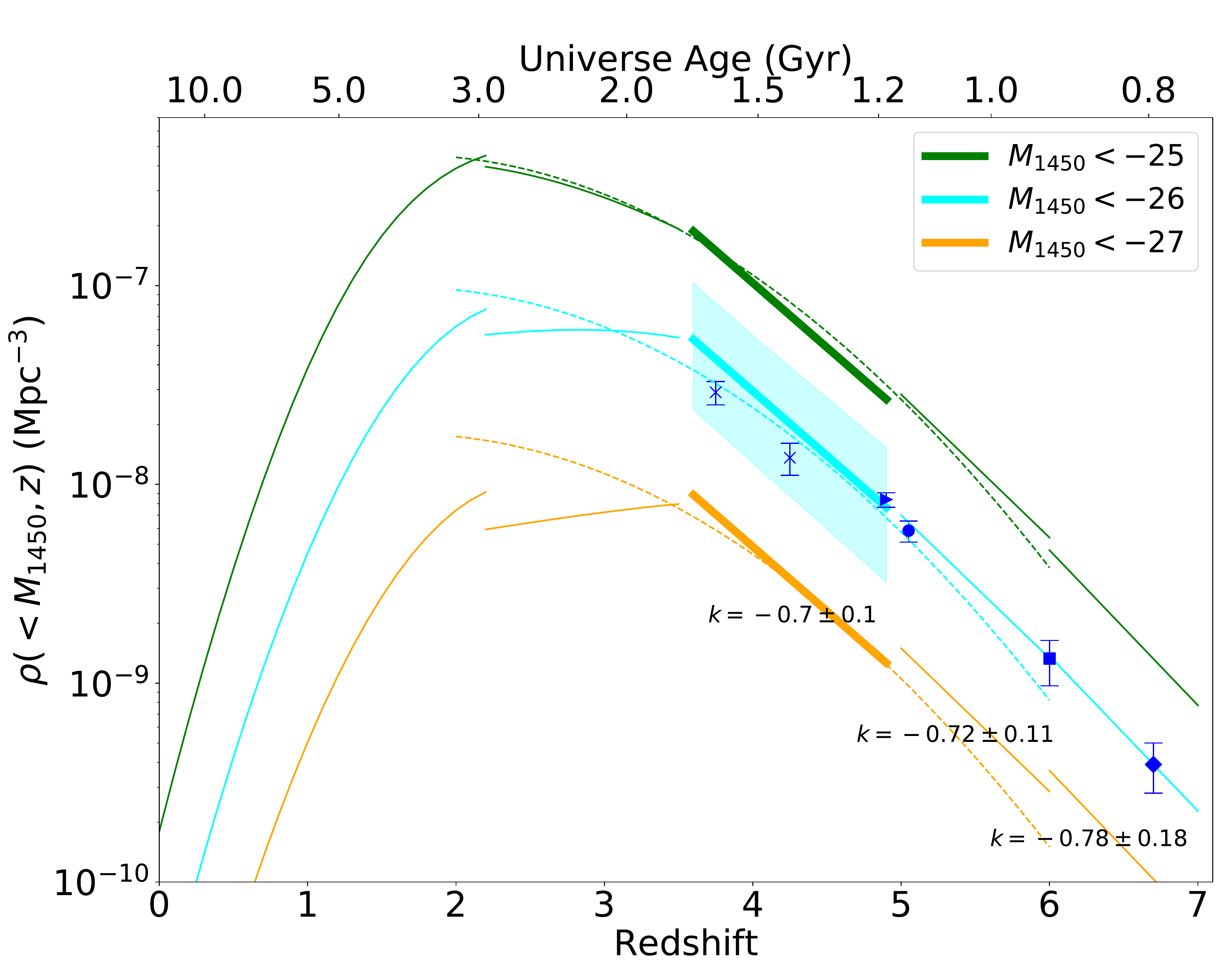}
\caption{Cumulative density evolution of quasars at $0<z<7$. The green, cyan, orange lines represent  magnitude ranges, $M_{1450}<-25,-26,-27$ mag, respectively. The bold solid lines are the densities at $3.5<z<5$ calculated from the PDE model in this paper. The cyan shaded region indicates the 1$\sigma$ uncertainty of our PDE model. The solid lines at the ranges of $0<z<3.5$, $5<z<6$, $6<z<7$ are from \citet{Ross_2013}, \citet{2016ApJ...833..222J}, and \citet{Wang_2019}, respectively. The dashed lines show the PDE model at $2<z<6$ from \citet{2021arXiv210306265K}. The blue symbols denote observational cumulative densities at $z=3.75, 4.25, 4.9, 5.05, 6.1$, and 6.7 measured by \citet{Richards_2006} (cross), \citet{2013ApJ...768..105M} (triangle), \citet{2016ApJ...829...33Y} (circle), \citet{2016ApJ...833..222J} (square), and \citet{Wang_2019} (diamond).}    \label{fig:density}
\end{figure}

The models of Cases 1, 2, and 3 have similar fitting performance in terms of $\chi^2_\nu$. The model in Case 4, with up to 8 free parameters, has likely overfitted the observed data because its $\chi_\nu^2$ is much smaller than 1. Besides, the PDE model in Case 1 has the smallest number of free parameters, and is enough to describe the evolution of the QLF at $z\sim3.5-5$. This is consistent with the conclusion of \citet{2021arXiv210306265K}.

We compare the four models in Figure \ref{fig:qlf_evol}. The Case 1 result (purple solid line) and the Case 4 result (purple dashed line) have similar fitting performance, so it is difficult to find the best-fit model based on current data. These two results are both higher than the result of \citet{2019MNRAS.488.1035K} in the bright end. This discrepancy can be attributed to the different data that two studies used. For example, \citet{2019MNRAS.488.1035K} did not use the data from \citet{Boutsia_2018} and \citet{Schindler_2019}. These data may increase the measurement of the bright-end QLF. In addition, the redshift coverage of \citet{2019MNRAS.488.1035K} is much larger than this work. The data in the range of $z<3.5$ and $z>5$ will affect the result in $3.5<z<5$ when computing the QLF evolution.

Compared with \citet{2021arXiv210306265K}, our QLF slopes are steeper. The measurements of the slopes sensitively depend on the data points at the brightest and faintest ends that usually suffer from large incompleteness and uncertainties. In addition, the combination of different samples introduces extra uncertainties that are often difficult to characterize. A large sample with a full coverage of the both ends is needed to improve the measurement of the QLF.

Finally, we explore the cumulative spatial density evolution of quasars at high redshift. The spatial density of quasars brighter than a given magnitude $M$ is calculated by integrating the QLF,
\begin{equation} \label{eq:density}
\rho(<M,z)=\int_{-\infty}^{M}\Phi(M,z)dM,
\end{equation}
where we use the PDE model (case 1) of this work as $\Phi(M,z)$. Figure \ref{fig:density} shows the cumulative density as a function of redshift for different magnitude ranges using our model and previous results \citep{Richards_2006,Ross_2013,2013ApJ...768..105M,2016ApJ...833..222J,2016ApJ...829...33Y,Wang_2019,2021arXiv210306265K}. It indicates a rapid density decline from $z\sim3.5$ to $z\sim7$, consistent with previous studies. For example, \citet{2001AJ....121...54F} fit an exponential decline, $\rho(<M,z)\propto10^{kz}$, and found $k=-0.47$ at high redshift. \citet{2013ApJ...768..105M} found that the slope at $4<z<5$ was  $k=-0.38\pm0.07$. Previous results also suggested that the spatial density of quasars drops faster with increasing redshift at $z>3.5$. As shown in Figure \ref{fig:density}, the slopes at $5<z<6$ and $6<z<7$ are $k=-0.72\pm0.11$ and $k=-0.78\pm0.18$, respectively \citep{2016ApJ...833..222J,Wang_2019}. The PDE model from \citet{2021arXiv210306265K} (dashed lines) also supports this scenario as shown in Figure \ref{fig:density}.

From our sample, we measured $k=-0.7\pm0.1$ at $3.5<z<5$, which is slightly steeper than previous measurements for the same redshift range, but is similar to those measured at $z=5\sim7$. The main reason for the steeper slope is that our binned QLF within $-26.5<M_{1450}<-25.5$ at $z=3.8$ is about 1.5 times the previous results (see grey squares in Figure \ref{fig:binned_qlf_compare} and blue crosses in Figure \ref{fig:density}). However, the cumulative densities between our model and the observed results are consistent within 1$\sigma$. To confirm our results, we need a larger and more complete sample, such as a quasar sample from the Chinese Space Station Telescope wide-area slitless spectroscopic survey \citep{Zhan_2021}. If confirmed, the quasar density at $3.5<z<5$ declines faster than previous measurements, and as fast as the density evolution at $z>5$.

\section{Summary} \label{sec:sum}

In this paper we have built a sample of more than 1000 quasars at $z>3$, including 974 quasars in 1292 deg$^2$ of the SDSS overlap regions and 224 quasars in 225 deg$^2$ of Stripe 82. The spectroscopic observations were conducted by the SDSS-III BOSS. The sample spans an absolute magnitude range of $-27.5<M_{1450}<-24.0$ mag. This is roughly 1.5 mag fainter than the SDSS main quasar sample selected from the single-epoch data. 

We have constructed QLFs at $3.5<z<5$ based on this sample and studied quasar evolution from $z=5$ to 3.5. We first corrected sample incompleteness caused by the misclassification of the object morphology, the color selection of the candidates, and the incomplete spectroscopy of the candidates. We then derived the binned QLFs at $3.6<z<4.0$, $4.0<z<4.5$, and $4.5<z<4.9$, and modeled the QLFs using a double-power law form. The luminosity coverage of our sample is not large enough to constrain all parameters in the double-power law model, so we fixed the bright-end slope $\beta$. We found that the faint-end slopes for the three redshift ranges are $\alpha\sim-1.8$, with moderate to large uncertainties from $0.1-0.3$ to $>0.5$. The relatively large uncertainties are mainly due to the relatively small sample size and the fact that our sample does not reach a very low luminosity.

We have made use of some studies from the literature and improved the measurement of the QLFs. We combined their binned QLFs with ours and characterized the QLFs in a larger luminosity range of $-29<M_{1450}<-23$ mag. We found that the faint-end slopes of the QLFs are around $-1.7$ and $-1.6$ and the bright-end slopes are from $-4.0$ to $-3.2$. Finally, we investigated the evolution of the QLFs from $z\sim5$ to 3.5 and found that a simple PDE model can efficiently describe the QLF evolution in this redshift range. This is consistent with some recent results. We also found that the quasar density at $3.5<z<5$ declines faster than previously thought, and its evolution parameter $k$ is similar to that at $z>5$.

\acknowledgments
We acknowledge support from the National Key R\&D Program of China (2016YFA0400703), the National Science Foundation of China (11721303, 11890693), and the science research grants from the China Manned Space Project with NO. CMS-CSST-2021-A05. 

Funding for SDSS-III has been provided by the Alfred P. Sloan Foundation, the Participating Institutions, the National Science Foundation, and the U.S. Department of Energy Office of Science. The SDSS-III web site is http://www.sdss3.org/. SDSS-III is managed by the Astrophysical Research Consortium for the Participating Institutions of the SDSS-III Collaboration including the University of Arizona, the Brazilian Participation Group, Brookhaven National Laboratory, Carnegie Mellon University, University of Florida, the French Participation Group, the German Participation Group, Harvard University, the Instituto de Astrofisica de Canarias, the Michigan State/Notre Dame/JINA Participation Group, Johns Hopkins University, Lawrence Berkeley National Laboratory, Max Planck Institute for Astrophysics, Max Planck Institute for Extraterrestrial Physics, New Mexico State University, New York University, Ohio State University, Pennsylvania State University, University of Portsmouth, Princeton University, the Spanish Participation Group, University of Tokyo, University of Utah, Vanderbilt University, University of Virginia, University of Washington, and Yale University.

\facilities{SDSS}

\software{ASERA\citep{2013A&C.....3...65Y}, Astropy\citep{2013A&A...558A..33A,2018AJ....156..123A}, emcee\citep{2013PASP..125..306F}, Topcat\citep{2005ASPC..347...29T}}

\newpage
\appendix
\section{Selection Criteria} \label{App:sel}
The criteria for the $gri$ candidates in overlap regions at $19.0<i<20.2$ mag are as follows,
\begin{equation} \label{eq:overlap_gri_1}
	\left\{ 
		\begin{array}{ll}
   	i_{\rm{err}}<0.06 \\
   	|i_1-i_2|<0.25 \\
   	u>21.5 \\
   	g>20.0, \ {\rm or} \ u-g>1.5 \\
   	g-r>0.65 \\
   	g-r>2.0, \ {\rm or} \ r-i<0.44\,(g-r)-0.286 \\
		19.0 < i < 20.2 \\
   	-1.0<r-i<1.0 \\
   	-1.0<i-z<1.0.
      \end{array}
   \right.
\end{equation}
The criteria for the $gri$ candidates in overlap regions at $20.2<i<20.8$ mag are as follows,
\begin{equation} \label{eq:overlap_gri_2}
	\left\{ 
		\begin{array}{ll}
   	i_{\rm{err}}<0.08 \\
   	|i_1-i_2|<0.30 \\
   	u>22.5 \\
   	g>21.0, \ {\rm or} \ u-g>1.5 \\
   	g-r>0.65 \\
   	g-r>2.1, \ {\rm or} \ r-i<0.44\,(g-r)-0.336 \\
		20.2 < i < 20.8 \\
   	-1.0<r-i<0.9 \\
   	-1.0<i-z<1.0.
      \end{array}
   \right.
\end{equation}
The criteria for the $gri$ candidates in overlap regions at $20.8<i<21.3$ mag are as follows,
\begin{equation} \label{eq:overlap_gri_3}
	\left\{ 
		\begin{array}{ll}
   	i_{\rm{err}}<0.10 \\
   	|i_1-i_2|<0.35 \\
   	u>23.0 \\
   	g>21.5, \ {\rm or} \ u-g>1.5 \\
   	g-r>0.65 \\
   	r-i<0.44\,(g-r)-0.436 \\
		20.8 < i < 21.3 \\
   	-1.0<r-i<0.8 \\
   	-1.0<i-z<1.0.
      \end{array}
   \right.
\end{equation}
The criteria for the $riz$ candidates in overlap regions at $19.0<i<20.2$ mag are as follows,
\begin{equation} \label{eq:overlap_riz_1}
   \left\{ 
      \begin{array}{ll}
    i_{\rm{err}}<0.06 \\
   	|i_1-i_2|<0.25 \\
   	u>22.0 \\
    g>21.5 \\
    r-i>0.6 \\
    -1.0<i-z<0.52(r-i)-0.312 \\
	  19.0 < i < 20.2. \\
      \end{array}
   \right.
\end{equation}
The criteria for the $riz$ candidates in overlap regions at $20.2<i<20.8$ mag are as follows,
\begin{equation} \label{eq:overlap_riz_2}
   \left\{ 
      \begin{array}{ll}
    i_{\rm{err}}<0.08 \\
   	|i_1-i_2|<0.30 \\
   	u>23.0 \\
    g>22.5 \\
    r-i>0.6 \\
    -1.0<i-z<0.52(r-i)-0.412 \\
	  20.2 < i < 20.8. \\
      \end{array}
   \right.
\end{equation}
The criteria for the $riz$ candidates in overlap regions at $20.8<i<21.3$ mag are as follows,
\begin{equation} \label{eq:overlap_riz_3}
   \left\{ 
      \begin{array}{ll}
    i_{\rm{err}}<0.10 \\
   	|i_1-i_2|<0.35 \\
   	u>23.5 \\
    g>23.0 \\
    r-i>0.8 \\
    -1.0<i-z<0.52(r-i)-0.812 \\
	  20.8 < i < 21.3. \\
      \end{array}
   \right.
\end{equation}
The criteria for the $gri$ candidates in Stripe 82 at $i<20.2$ mag are as follows,
\begin{equation} \label{eq:82_gri_1}
	\left\{ 
		\begin{array}{ll}
   	u>21.5 \\
   	g>20.0, \ {\rm or} \ u-g>1.5 \\
   	g-r>0.65 \\
   	g-r>2.0, \ {\rm or} \ r-i<0.44\,(g-r)-0.286 \\
		i < 20.2 \\
   	-1.0<r-i<1.0 \\
   	-1.0<i-z<1.0.
      \end{array}
   \right.
\end{equation}
The criteria for the $gri$ candidates in Stripe 82 at $20.2<i<20.8$ mag are as follows,
\begin{equation} \label{eq:82_gri_2}
	\left\{ 
		\begin{array}{ll}
   	u>22.5 \\
   	g>21.0, \ {\rm or} \ u-g>1.5 \\
   	g-r>0.65 \\
   	g-r>2.1, \ {\rm or} \ r-i<0.44\,(g-r)-0.336 \\
		20.2 < i < 20.8 \\
   	-1.0<r-i<0.9 \\
   	-1.0<i-z<1.0.
      \end{array}
   \right.
\end{equation}
The criteria for the $gri$ candidates in Stripe 82 at $20.8<i<21.5$ mag are as follows,
\begin{equation} \label{eq:82_gri_3}
	\left\{ 
		\begin{array}{ll}
   	u>23.0 \\
   	g>21.5, \ {\rm or} \ u-g>1.5 \\
   	g-r>0.65 \\
   	r-i<0.44\,(g-r)-0.436 \\
		20.8 < i < 21.3 \\
   	-1.0<r-i<0.8 \\
   	-1.0<i-z<1.0.
      \end{array}
   \right.
\end{equation}
The criteria for the $riz$ candidates in Stripe 82 at $i<20.2$ mag are as follows,
\begin{equation} \label{eq:82_riz_1}
   \left\{ 
      \begin{array}{ll}
   	u>22.0 \\
    g>21.5 \\
    r-i>0.6 \\
    -1.0<i-z<0.52(r-i)-0.262 \\
	  i < 20.2. \\
      \end{array}
   \right.
\end{equation}
The criteria for the $riz$ candidates in Stripe 82 at $20.2<i<20.8$ mag are as follows,
\begin{equation} \label{eq:82_riz_2}
   \left\{ 
      \begin{array}{ll}
   	u>23.0 \\
    g>22.5 \\
    r-i>0.6 \\
    -1.0<i-z<0.52(r-i)-0.362 \\
	  20.2 < i < 20.8. \\
      \end{array}
   \right.
\end{equation}
The criteria for the $riz$ candidates in Stripe 82 at $20.8<i<21.5$ mag are as follows,
\begin{equation} \label{eq:82_riz_3}
   \left\{ 
      \begin{array}{ll}
   	u>23.5 \\
    g>23.0 \\
    r-i>0.6 \\
    -1.0<i-z<0.52(r-i)-0.462 \\
	  20.8 < i < 21.5. \\
      \end{array}
   \right.
\end{equation}

\section{Maximum Likelihood Estimation} \label{App:MLE}
\subsection{Two Contributions to The Uncertainty} 
When we use the maximum likelihood estimation to fit the binned QLFs, these data have the uncertainties of both y-axis and x-axis. The uncertainty of y-axis ($\sigma_{\mathrm{obs}}$) is directly the uncertainty of the observed binned QLFs ($\Phi_{\mathrm{obs}}$). The uncertainty of x-axis ($\sigma_{\mathrm{mag}}$) is taken from the magnitude bin size. We can estimate such effect on the fitting through the error propagation:
\begin{equation} 
\sigma_{\mathrm{extra}}=0.4\sigma_{\mathrm{mag}}\Phi_{\mathrm{obs}}\frac{(\alpha+1)\mathrm{log}10\times 10^{0.4(\alpha+1)(M-M^*)}+(\beta+1)\mathrm{log}10\times 10^{0.4(\beta+1)(M-M^*)}}{10^{0.4(\alpha+1)(M-M^*)}+10^{0.4(\beta+1)(M-M^*)}},
\end{equation}
Then the ultimate uncertainty used in the likelihood is: $\sigma=\sqrt{\sigma_{\mathrm{obs}}^2+\sigma_{\mathrm{extra}}^2}$.

\subsection{The Chosen Priors on The Parameters} 
Actually, we choose the initial values without much consideration: $\mathrm{log}\Phi^*=-7.9$, $M^*=-26.5$, $\alpha=-1.3$, $\beta=-3.0$. We also constrain the ranges of parameters: $-9<\mathrm{log}\Phi^*<-6$, $-28<M^*<-24$, $-3<\alpha<0$, $-6<\beta<-2$.

\newpage
\bibliography{QLF}{}

\begin{thebibliography}{}
\expandafter\ifx\csname natexlab\endcsname\relax\def\natexlab#1{#1}\fi
\providecommand{\url}[1]{\href{#1}{#1}}
\providecommand{\dodoi}[1]{doi:~\href{http://doi.org/#1}{\nolinkurl{#1}}}
\providecommand{\doeprint}[1]{\href{http://ascl.net/#1}{\nolinkurl{http://ascl.net/#1}}}
\providecommand{\doarXiv}[1]{\href{https://arxiv.org/abs/#1}{\nolinkurl{https://arxiv.org/abs/#1}}}

\bibitem[{{Akiyama} {et~al.}(2018){Akiyama}, {He}, {Ikeda}, {Niida}, {Nagao},
  {Bosch}, {Coupon}, {Enoki}, {Imanishi}, {Kashikawa}, {Kawaguchi}, {Komiyama},
  {Lee}, {Matsuoka}, {Miyazaki}, {Nishizawa}, {Oguri}, {Ono}, {Onoue}, {Ouchi},
  {Schulze}, {Silverman}, {Tanaka}, {Tanaka}, {Terashima}, {Toba}, \&
  {Ueda}}]{2018PASJ...70S..34A}
{Akiyama}, M., {He}, W., {Ikeda}, H., {et~al.} 2018, \pasj, 70, S34,
  \dodoi{10.1093/pasj/psx091}

\bibitem[{{Alam} {et~al.}(2015){Alam}, {Albareti}, {Allende Prieto}, {Anders},
  {Anderson}, {Anderton}, {Andrews}, {Armengaud}, {Aubourg}, {Bailey}, {Basu},
  {Bautista}, {Beaton}, {Beers}, {Bender}, {Berlind}, {Beutler}, {Bhardwaj},
  {Bird}, {Bizyaev}, {Blake}, {Blanton}, {Blomqvist}, {Bochanski}, {Bolton},
  {Bovy}, {Shelden Bradley}, {Brandt}, {Brauer}, {Brinkmann}, {Brown},
  {Brownstein}, {Burden}, {Burtin}, {Busca}, {Cai}, {Capozzi}, {Carnero
  Rosell}, {Carr}, {Carrera}, {Chambers}, {Chaplin}, {Chen}, {Chiappini},
  {Chojnowski}, {Chuang}, {Clerc}, {Comparat}, {Covey}, {Croft}, {Cuesta},
  {Cunha}, {da Costa}, {Da Rio}, {Davenport}, {Dawson}, {De Lee}, {Delubac},
  {Deshpande}, {Dhital}, {Dutra-Ferreira}, {Dwelly}, {Ealet}, {Ebelke},
  {Edmondson}, {Eisenstein}, {Ellsworth}, {Elsworth}, {Epstein}, {Eracleous},
  {Escoffier}, {Esposito}, {Evans}, {Fan}, {Fern{\'a}ndez-Alvar}, {Feuillet},
  {Filiz Ak}, {Finley}, {Finoguenov}, {Flaherty}, {Fleming}, {Font-Ribera},
  {Foster}, {Frinchaboy}, {Galbraith-Frew}, {Garc{\'\i}a},
  {Garc{\'\i}a-Hern{\'a}ndez}, {Garc{\'\i}a P{\'e}rez}, {Gaulme}, {Ge},
  {G{\'e}nova-Santos}, {Georgakakis}, {Ghezzi}, {Gillespie}, {Girardi},
  {Goddard}, {Gontcho}, {Gonz{\'a}lez Hern{\'a}ndez}, {Grebel}, {Green},
  {Grieb}, {Grieves}, {Gunn}, {Guo}, {Harding}, {Hasselquist}, {Hawley},
  {Hayden}, {Hearty}, {Hekker}, {Ho}, {Hogg}, {Holley-Bockelmann}, {Holtzman},
  {Honscheid}, {Huber}, {Huehnerhoff}, {Ivans}, {Jiang}, {Johnson},
  {Kinemuchi}, {Kirkby}, {Kitaura}, {Klaene}, {Knapp}, {Kneib}, {Koenig},
  {Lam}, {Lan}, {Lang}, {Laurent}, {Le Goff}, {Leauthaud}, {Lee}, {Lee},
  {Licquia}, {Liu}, {Long}, {L{\'o}pez-Corredoira}, {Lorenzo-Oliveira},
  {Lucatello}, {Lundgren}, {Lupton}, {Mack}, {Mahadevan}, {Maia}, {Majewski},
  {Malanushenko}, {Malanushenko}, {Manchado}, {Manera}, {Mao}, {Maraston},
  {Marchwinski}, {Margala}, {Martell}, {Martig}, {Masters}, {Mathur},
  {McBride}, {McGehee}, {McGreer}, {McMahon}, {M{\'e}nard}, {Menzel},
  {Merloni}, {M{\'e}sz{\'a}ros}, {Miller}, {Miralda-Escud{\'e}}, {Miyatake},
  {Montero-Dorta}, {More}, {Morganson}, {Morice-Atkinson}, {Morrison},
  {Mosser}, {Muna}, {Myers}, {Nandra}, {Newman}, {Neyrinck}, {Nguyen},
  {Nichol}, {Nidever}, {Noterdaeme}, {Nuza}, {O'Connell}, {O'Connell},
  {O'Connell}, {Ogando}, {Olmstead}, {Oravetz}, {Oravetz}, {Osumi}, {Owen},
  {Padgett}, {Padmanabhan}, {Paegert}, {Palanque-Delabrouille}, {Pan},
  {Parejko}, {P{\^a}ris}, {Park}, {Pattarakijwanich}, {Pellejero-Ibanez},
  {Pepper}, \& {Percival}}]{2015ApJS..219...12A}
{Alam}, S., {Albareti}, F.~D., {Allende Prieto}, C., {et~al.} 2015, \apjs, 219,
  12, \dodoi{10.1088/0067-0049/219/1/12}

\bibitem[{{Annis} {et~al.}(2014){Annis}, {Soares-Santos}, {Strauss}, {Becker},
  {Dodelson}, {Fan}, {Gunn}, {Hao}, {Ivezi{\'c}}, {Jester}, {Jiang},
  {Johnston}, {Kubo}, {Lampeitl}, {Lin}, {Lupton}, {Miknaitis}, {Seo}, {Simet},
  \& {Yanny}}]{2014ApJ...794..120A}
{Annis}, J., {Soares-Santos}, M., {Strauss}, M.~A., {et~al.} 2014, \apj, 794,
  120, \dodoi{10.1088/0004-637X/794/2/120}

\bibitem[{{Astropy Collaboration} {et~al.}(2013){Astropy Collaboration},
  {Robitaille}, {Tollerud}, {Greenfield}, {Droettboom}, {Bray}, {Aldcroft},
  {Davis}, {Ginsburg}, {Price-Whelan}, {Kerzendorf}, {Conley}, {Crighton},
  {Barbary}, {Muna}, {Ferguson}, {Grollier}, {Parikh}, {Nair}, {Unther},
  {Deil}, {Woillez}, {Conseil}, {Kramer}, {Turner}, {Singer}, {Fox}, {Weaver},
  {Zabalza}, {Edwards}, {Azalee Bostroem}, {Burke}, {Casey}, {Crawford},
  {Dencheva}, {Ely}, {Jenness}, {Labrie}, {Lim}, {Pierfederici}, {Pontzen},
  {Ptak}, {Refsdal}, {Servillat}, \& {Streicher}}]{2013A&A...558A..33A}
{Astropy Collaboration}, {Robitaille}, T.~P., {Tollerud}, E.~J., {et~al.} 2013,
  \aap, 558, A33, \dodoi{10.1051/0004-6361/201322068}

\bibitem[{{Astropy Collaboration} {et~al.}(2018){Astropy Collaboration},
  {Price-Whelan}, {Sip{\H{o}}cz}, {G{\"u}nther}, {Lim}, {Crawford}, {Conseil},
  {Shupe}, {Craig}, {Dencheva}, {Ginsburg}, {VanderPlas}, {Bradley},
  {P{\'e}rez-Su{\'a}rez}, {de Val-Borro}, {Aldcroft}, {Cruz}, {Robitaille},
  {Tollerud}, {Ardelean}, {Babej}, {Bach}, {Bachetti}, {Bakanov}, {Bamford},
  {Barentsen}, {Barmby}, {Baumbach}, {Berry}, {Biscani}, {Boquien}, {Bostroem},
  {Bouma}, {Brammer}, {Bray}, {Breytenbach}, {Buddelmeijer}, {Burke},
  {Calderone}, {Cano Rodr{\'\i}guez}, {Cara}, {Cardoso}, {Cheedella}, {Copin},
  {Corrales}, {Crichton}, {D'Avella}, {Deil}, {Depagne}, {Dietrich}, {Donath},
  {Droettboom}, {Earl}, {Erben}, {Fabbro}, {Ferreira}, {Finethy}, {Fox},
  {Garrison}, {Gibbons}, {Goldstein}, {Gommers}, {Greco}, {Greenfield},
  {Groener}, {Grollier}, {Hagen}, {Hirst}, {Homeier}, {Horton}, {Hosseinzadeh},
  {Hu}, {Hunkeler}, {Ivezi{\'c}}, {Jain}, {Jenness}, {Kanarek}, {Kendrew},
  {Kern}, {Kerzendorf}, {Khvalko}, {King}, {Kirkby}, {Kulkarni}, {Kumar},
  {Lee}, {Lenz}, {Littlefair}, {Ma}, {Macleod}, {Mastropietro}, {McCully},
  {Montagnac}, {Morris}, {Mueller}, {Mumford}, {Muna}, {Murphy}, {Nelson},
  {Nguyen}, {Ninan}, {N{\"o}the}, {Ogaz}, {Oh}, {Parejko}, {Parley}, {Pascual},
  {Patil}, {Patil}, {Plunkett}, {Prochaska}, {Rastogi}, {Reddy Janga},
  {Sabater}, {Sakurikar}, {Seifert}, {Sherbert}, {Sherwood-Taylor}, {Shih},
  {Sick}, {Silbiger}, {Singanamalla}, {Singer}, {Sladen}, {Sooley},
  {Sornarajah}, {Streicher}, {Teuben}, {Thomas}, {Tremblay}, {Turner},
  {Terr{\'o}n}, {van Kerkwijk}, {de la Vega}, {Watkins}, {Weaver}, {Whitmore},
  {Woillez}, {Zabalza}, \& {Astropy Contributors}}]{2018AJ....156..123A}
{Astropy Collaboration}, {Price-Whelan}, A.~M., {Sip{\H{o}}cz}, B.~M., {et~al.}
  2018, \aj, 156, 123, \dodoi{10.3847/1538-3881/aabc4f}

\bibitem[{{Avni} \& {Bahcall}(1980)}]{1980ApJ...235..694A}
{Avni}, Y., \& {Bahcall}, J.~N. 1980, \apj, 235, 694, \dodoi{10.1086/157673}

\bibitem[{{Bongiorno} {et~al.}(2007){Bongiorno}, {Zamorani}, {Gavignaud},
  {Marano}, {Paltani}, {Mathez}, {M{\o}ller}, {Picat}, {Cirasuolo},
  {Lamareille}, {Bottini}, {Garilli}, {Le Brun}, {Le F{\`e}vre}, {Maccagni},
  {Scaramella}, {Scodeggio}, {Tresse}, {Vettolani}, {Zanichelli}, {Adami},
  {Arnouts}, {Bardelli}, {Bolzonella}, {Cappi}, {Charlot}, {Ciliegi},
  {Contini}, {Foucaud}, {Franzetti}, {Guzzo}, {Ilbert}, {Iovino}, {McCracken},
  {Marinoni}, {Mazure}, {Meneux}, {Merighi}, {Pell{\`o}}, {Pollo}, {Pozzetti},
  {Radovich}, {Zucca}, {Hatziminaoglou}, {Polletta}, {Bondi}, {Brinchmann},
  {Cucciati}, {de la Torre}, {Gregorini}, {Mellier}, {Merluzzi}, {Temporin},
  {Vergani}, \& {Walcher}}]{2007AA...472..443B}
{Bongiorno}, A., {Zamorani}, G., {Gavignaud}, I., {et~al.} 2007, \aap, 472,
  443, \dodoi{10.1051/0004-6361:20077611}

\bibitem[{Boutsia {et~al.}(2018)Boutsia, Grazian, Giallongo, Fiore, \&
  Civano}]{Boutsia_2018}
Boutsia, K., Grazian, A., Giallongo, E., Fiore, F., \& Civano, F. 2018, The
  Astrophysical Journal, 869, 20, \dodoi{10.3847/1538-4357/aae6c7}

\bibitem[{{Boutsia} {et~al.}(2021){Boutsia}, {Grazian}, {Fontanot},
  {Giallongo}, {Menci}, {Calderone}, {Cristiani}, {D'Odorico}, {Cupani},
  {Guarneri}, \& {Omizzolo}}]{2021arXiv210310446B}
{Boutsia}, K., {Grazian}, A., {Fontanot}, F., {et~al.} 2021, arXiv e-prints,
  arXiv:2103.10446.
\newblock \doarXiv{2103.10446}

\bibitem[{Bovy {et~al.}(2011)Bovy, Hennawi, Hogg, Myers, Kirkpatrick, Schlegel,
  Ross, Sheldon, McGreer, Schneider, \& Weaver}]{Bovy_2011}
Bovy, J., Hennawi, J.~F., Hogg, D.~W., {et~al.} 2011, The Astrophysical
  Journal, 729, 141, \dodoi{10.1088/0004-637x/729/2/141}

\bibitem[{{Boyle} {et~al.}(2000){Boyle}, {Shanks}, {Croom}, {Smith}, {Miller},
  {Loaring}, \& {Heymans}}]{2000MNRAS.317.1014B}
{Boyle}, B.~J., {Shanks}, T., {Croom}, S.~M., {et~al.} 2000, \mnras, 317, 1014,
  \dodoi{10.1046/j.1365-8711.2000.03730.x}

\bibitem[{{Boyle} {et~al.}(1988){Boyle}, {Shanks}, \&
  {Peterson}}]{1988MNRAS.235..935B}
{Boyle}, B.~J., {Shanks}, T., \& {Peterson}, B.~A. 1988, \mnras, 235, 935,
  \dodoi{10.1093/mnras/235.3.935}

\bibitem[{{Croom} {et~al.}(2004){Croom}, {Smith}, {Boyle}, {Shanks}, {Miller},
  {Outram}, \& {Loaring}}]{2004MNRAS.349.1397C}
{Croom}, S.~M., {Smith}, R.~J., {Boyle}, B.~J., {et~al.} 2004, \mnras, 349,
  1397, \dodoi{10.1111/j.1365-2966.2004.07619.x}

\bibitem[{{Dawson} {et~al.}(2013){Dawson}, {Schlegel}, {Ahn}, {Anderson},
  {Aubourg}, {Bailey}, {Barkhouser}, {Bautista}, {Beifiori}, {Berlind},
  {Bhardwaj}, {Bizyaev}, {Blake}, {Blanton}, {Blomqvist}, {Bolton}, {Borde},
  {Bovy}, {Brandt}, {Brewington}, {Brinkmann}, {Brown}, {Brownstein}, {Bundy},
  {Busca}, {Carithers}, {Carnero}, {Carr}, {Chen}, {Comparat}, {Connolly},
  {Cope}, {Croft}, {Cuesta}, {da Costa}, {Davenport}, {Delubac}, {de Putter},
  {Dhital}, {Ealet}, {Ebelke}, {Eisenstein}, {Escoffier}, {Fan}, {Filiz Ak},
  {Finley}, {Font-Ribera}, {G{\'e}nova-Santos}, {Gunn}, {Guo}, {Haggard},
  {Hall}, {Hamilton}, {Harris}, {Harris}, {Ho}, {Hogg}, {Holder}, {Honscheid},
  {Huehnerhoff}, {Jordan}, {Jordan}, {Kauffmann}, {Kazin}, {Kirkby}, {Klaene},
  {Kneib}, {Le Goff}, {Lee}, {Long}, {Loomis}, {Lundgren}, {Lupton}, {Maia},
  {Makler}, {Malanushenko}, {Malanushenko}, {Mandelbaum}, {Manera}, {Maraston},
  {Margala}, {Masters}, {McBride}, {McDonald}, {McGreer}, {McMahon}, {Mena},
  {Miralda-Escud{\'e}}, {Montero-Dorta}, {Montesano}, {Muna}, {Myers},
  {Naugle}, {Nichol}, {Noterdaeme}, {Nuza}, {Olmstead}, {Oravetz}, {Oravetz},
  {Owen}, {Padmanabhan}, {Palanque-Delabrouille}, {Pan}, {Parejko},
  {P{\^a}ris}, {Percival}, {P{\'e}rez-Fournon}, {P{\'e}rez-R{\`a}fols},
  {Petitjean}, {Pfaffenberger}, {Pforr}, {Pieri}, {Prada}, {Price-Whelan},
  {Raddick}, {Rebolo}, {Rich}, {Richards}, {Rockosi}, {Roe}, {Ross}, {Ross},
  {Rossi}, {Rubi{\~n}o-Martin}, {Samushia}, {S{\'a}nchez}, {Sayres}, {Schmidt},
  {Schneider}, {Sc{\'o}ccola}, {Seo}, {Shelden}, {Sheldon}, {Shen}, {Shu},
  {Slosar}, {Smee}, {Snedden}, {Stauffer}, {Steele}, {Strauss}, {Streblyanska},
  {Suzuki}, {Swanson}, {Tal}, {Tanaka}, {Thomas}, {Tinker}, {Tojeiro},
  {Tremonti}, {Vargas Maga{\~n}a}, {Verde}, {Viel}, {Wake}, {Watson}, {Weaver},
  {Weinberg}, {Weiner}, {West}, {White}, {Wood-Vasey}, {Yeche}, {Zehavi},
  {Zhao}, \& {Zheng}}]{2013AJ....145...10D}
{Dawson}, K.~S., {Schlegel}, D.~J., {Ahn}, C.~P., {et~al.} 2013, \aj, 145, 10,
  \dodoi{10.1088/0004-6256/145/1/10}

\bibitem[{{Eisenstein} {et~al.}(2011){Eisenstein}, {Weinberg}, {Agol},
  {Aihara}, {Allende Prieto}, {Anderson}, {Arns}, {Aubourg}, {Bailey},
  {Balbinot}, {Barkhouser}, {Beers}, {Berlind}, {Bickerton}, {Bizyaev},
  {Blanton}, {Bochanski}, {Bolton}, {Bosman}, {Bovy}, {Brandt}, {Breslauer},
  {Brewington}, {Brinkmann}, {Brown}, {Brownstein}, {Burger}, {Busca},
  {Campbell}, {Cargile}, {Carithers}, {Carlberg}, {Carr}, {Chang}, {Chen},
  {Chiappini}, {Comparat}, {Connolly}, {Cortes}, {Croft}, {Cunha}, {da Costa},
  {Davenport}, {Dawson}, {De Lee}, {Porto de Mello}, {de Simoni}, {Dean},
  {Dhital}, {Ealet}, {Ebelke}, {Edmondson}, {Eiting}, {Escoffier}, {Esposito},
  {Evans}, {Fan}, {Femen{\'\i}a Castell{\'a}}, {Dutra Ferreira}, {Fitzgerald},
  {Fleming}, {Font-Ribera}, {Ford}, {Frinchaboy}, {Garc{\'\i}a P{\'e}rez},
  {Gaudi}, {Ge}, {Ghezzi}, {Gillespie}, {Gilmore}, {Girardi}, {Gott}, {Gould},
  {Grebel}, {Gunn}, {Hamilton}, {Harding}, {Harris}, {Hawley}, {Hearty},
  {Hennawi}, {Gonz{\'a}lez Hern{\'a}ndez}, {Ho}, {Hogg}, {Holtzman},
  {Honscheid}, {Inada}, {Ivans}, {Jiang}, {Jiang}, {Johnson}, {Jordan},
  {Jordan}, {Kauffmann}, {Kazin}, {Kirkby}, {Klaene}, {Knapp}, {Kneib},
  {Kochanek}, {Koesterke}, {Kollmeier}, {Kron}, {Lampeitl}, {Lang}, {Lawler},
  {Le Goff}, {Lee}, {Lee}, {Leisenring}, {Lin}, {Liu}, {Long}, {Loomis},
  {Lucatello}, {Lundgren}, {Lupton}, {Ma}, {Ma}, {MacDonald}, {Mack},
  {Mahadevan}, {Maia}, {Majewski}, {Makler}, {Malanushenko}, {Malanushenko},
  {Mand elbaum}, {Maraston}, {Margala}, {Maseman}, {Masters}, {McBride},
  {McDonald}, {McGreer}, {McMahon}, {Mena Requejo}, {M{\'e}nard},
  {Miralda-Escud{\'e}}, {Morrison}, {Mullally}, {Muna}, {Murayama}, {Myers},
  {Naugle}, {Neto}, {Nguyen}, {Nichol}, {Nidever}, {O'Connell}, {Ogando},
  {Olmstead}, {Oravetz}, {Padmanabhan}, {Paegert}, {Palanque-Delabrouille},
  {Pan}, {Pandey}, {Parejko}, {P{\^a}ris}, {Pellegrini}, {Pepper}, {Percival},
  {Petitjean}, {Pfaffenberger}, {Pforr}, {Phleps}, {Pichon}, {Pieri}, {Prada},
  {Price-Whelan}, {Raddick}, {Ramos}, {Reid}, {Reyle}, {Rich}, {Richards},
  {Rieke}, {Rieke}, {Rix}, {Robin}, {Rocha-Pinto}, {Rockosi}, {Roe},
  {Rollinde}, {Ross}, {Ross}, {Rossetto}, {S{\'a}nchez}, {Santiago}, {Sayres},
  {Schiavon}, {Schlegel}, {Schlesinger}, {Schmidt}, {Schneider}, {Sellgren},
  {Shelden}, {Sheldon}, {Shetrone}, {Shu}, {Silverman}, {Simmerer}, {Simmons},
  {Sivarani}, {Skrutskie}, {Slosar}, {Smee}, {Smith}, {Snedden}, {Stassun},
  {Steele}, {Steinmetz}, {Stockett}, {Stollberg}, {Strauss}, {Szalay},
  {Tanaka}, {Thakar}, {Thomas}, {Tinker}, {Tofflemire}, {Tojeiro}, {Tremonti},
  {Vargas Maga{\~n}a}, {Verde}, {Vogt}, {Wake}, {Wan}, {Wang}, {Weaver},
  {White}, {White}, {Wilson}, {Wisniewski}, {Wood-Vasey}, {Yanny}, {Yasuda},
  {Y{\`e}che}, {York}, {Young}, {Zasowski}, {Zehavi}, \&
  {Zhao}}]{2011AJ....142...72E}
{Eisenstein}, D.~J., {Weinberg}, D.~H., {Agol}, E., {et~al.} 2011, \aj, 142,
  72, \dodoi{10.1088/0004-6256/142/3/72}

\bibitem[{{Fan}(1999)}]{1999AJ....117.2528F}
{Fan}, X. 1999, \aj, 117, 2528, \dodoi{10.1086/300848}

\bibitem[{{Fan} {et~al.}(1999){Fan}, {Strauss}, {Schneider}, {Gunn}, {Lupton},
  {Yanny}, {Anderson}, {Anderson}, {Annis}, {Bahcall}, {Bakken}, {Bastian},
  {Berman}, {Boroski}, {Briegel}, {Briggs}, {Brinkmann}, {Carr}, {Colestock},
  {Connolly}, {Crocker}, {Csabai}, {Czarapata}, {Davis}, {Doi}, {Elms},
  {Evans}, {Federwitz}, {Frieman}, {Fukugita}, {Gurbani}, {Harris}, {Heckman},
  {Hennessy}, {Hindsley}, {Holmgren}, {Hull}, {Ichikawa}, {Ichikawa},
  {Ivezi{\'c} }, {Kent}, {Knapp}, {Kron}, {Lamb}, {Leger}, {Limmongkol},
  {Lindenmeyer}, {Long}, {Loveday}, {MacKinnon}, {Mannery}, {Mantsch},
  {Margon}, {McKay}, {Munn}, {Nash}, {Newberg}, {Nichol}, {Nicinski},
  {Okamura}, {Ostriker}, {Owen}, {Pauls}, {Peoples}, {Petravick}, {Pier},
  {Pordes}, {Prosapio}, {Rechenmacher}, {Richards}, {Richmond}, {Rivetta},
  {Rockosi}, {Sandford}, {Sergey}, {Sekiguchi}, {Shimasaku}, {Siegmund},
  {Smith}, {Stoughton}, {Szalay}, {Szokoly}, {Tucker}, {Vogeley}, {Waddell},
  {Wang}, {Weinberg}, {Yasuda}, \& {York}}]{1999AJ....118....1F}
{Fan}, X., {Strauss}, M.~A., {Schneider}, D.~P., {et~al.} 1999, \aj, 118, 1,
  \dodoi{10.1086/300944}

\bibitem[{{Fan} {et~al.}(2001{\natexlab{a}}){Fan}, {Strauss}, {Schneider},
  {Gunn}, {Lupton}, {Becker}, {Davis}, {Newman}, {Richards}, {White},
  {Anderson}, {Annis}, {Bahcall}, {Brunner}, {Csabai}, {Hennessy}, {Hindsley},
  {Fukugita}, {Kunszt}, {Ivezi{\'c}}, {Knapp}, {McKay}, {Munn}, {Pier},
  {Szalay}, \& {York}}]{2001AJ....121...54F}
---. 2001{\natexlab{a}}, \aj, 121, 54, \dodoi{10.1086/318033}

\bibitem[{{Fan} {et~al.}(2001{\natexlab{b}}){Fan}, {Narayanan}, {Lupton},
  {Strauss}, {Knapp}, {Becker}, {White}, {Pentericci}, {Leggett}, {Haiman},
  {Gunn}, {Ivezi{\'c}}, {Schneider}, {Anderson}, {Brinkmann}, {Bahcall},
  {Connolly}, {Csabai}, {Doi}, {Fukugita}, {Geballe}, {Grebel}, {Harbeck},
  {Hennessy}, {Lamb}, {Miknaitis}, {Munn}, {Nichol}, {Okamura}, {Pier},
  {Prada}, {Richards}, {Szalay}, \& {York}}]{2001AJ....122.2833F}
{Fan}, X., {Narayanan}, V.~K., {Lupton}, R.~H., {et~al.} 2001{\natexlab{b}},
  \aj, 122, 2833, \dodoi{10.1086/324111}

\bibitem[{{Fan} {et~al.}(2001{\natexlab{c}}){Fan}, {Strauss}, {Richards},
  {Newman}, {Becker}, {Schneider}, {Gunn}, {Davis}, {White}, {Lupton},
  {Anderson}, {Annis}, {Bahcall}, {Brunner}, {Csabai}, {Doi}, {Fukugita},
  {Hennessy}, {Hindsley}, {Ivezi{\'c}}, {Knapp}, {McKay}, {Munn}, {Pier},
  {Szalay}, \& {York}}]{2001AJ....121...31F}
{Fan}, X., {Strauss}, M.~A., {Richards}, G.~T., {et~al.} 2001{\natexlab{c}},
  \aj, 121, 31, \dodoi{10.1086/318032}

\bibitem[{{Fan} {et~al.}(2006){Fan}, {Strauss}, {Richards}, {Hennawi},
  {Becker}, {White}, {Diamond-Stanic}, {Donley}, {Jiang}, {Kim}, {Vestergaard},
  {Young}, {Gunn}, {Lupton}, {Knapp}, {Schneider}, {Brandt}, {Bahcall},
  {Barentine}, {Brinkmann}, {Brewington}, {Fukugita}, {Harvanek}, {Kleinman},
  {Krzesinski}, {Long}, {Neilsen}, {Nitta}, {Snedden}, \&
  {Voges}}]{2006AJ....131.1203F}
---. 2006, \aj, 131, 1203, \dodoi{10.1086/500296}

\bibitem[{Flesch(2021)}]{flesch2021million}
Flesch, E.~W. 2021, The Million Quasars (Milliquas) v7.2 Catalogue, now with
  VLASS associations. The inclusion of SDSS-DR16Q quasars is detailed.
\newblock \doarXiv{2105.12985}

\bibitem[{{Fontanot} {et~al.}(2007){Fontanot}, {Cristiani}, {Monaco}, {Nonino},
  {Vanzella}, {Brandt}, {Grazian}, \& {Mao}}]{2007AA...461...39F}
{Fontanot}, F., {Cristiani}, S., {Monaco}, P., {et~al.} 2007, \aap, 461, 39,
  \dodoi{10.1051/0004-6361:20066073}

\bibitem[{{Foreman-Mackey} {et~al.}(2013){Foreman-Mackey}, {Hogg}, {Lang}, \&
  {Goodman}}]{2013PASP..125..306F}
{Foreman-Mackey}, D., {Hogg}, D.~W., {Lang}, D., \& {Goodman}, J. 2013, \pasp,
  125, 306, \dodoi{10.1086/670067}

\bibitem[{{Fu} {et~al.}(2021){Fu}, {Wu}, {Yang}, {Brown}, {Feng}, {Ma}, \&
  {Li}}]{2021arXiv210209770F}
{Fu}, Y., {Wu}, X.-B., {Yang}, Q., {et~al.} 2021, arXiv e-prints,
  arXiv:2102.09770.
\newblock \doarXiv{2102.09770}

\bibitem[{{Gehrels}(1986)}]{1986ApJ...303..336G}
{Gehrels}, N. 1986, \apj, 303, 336, \dodoi{10.1086/164079}

\bibitem[{Glikman {et~al.}(2010)Glikman, Bogosavljevi{\'{c}}, Djorgovski,
  Stern, Dey, Jannuzi, \& Mahabal}]{Glikman_2010}
Glikman, E., Bogosavljevi{\'{c}}, M., Djorgovski, S.~G., {et~al.} 2010, The
  Astrophysical Journal, 710, 1498, \dodoi{10.1088/0004-637x/710/2/1498}

\bibitem[{Glikman {et~al.}(2011)Glikman, Djorgovski, Stern, Dey, Jannuzi, \&
  Lee}]{Glikman_2011}
Glikman, E., Djorgovski, S.~G., Stern, D., {et~al.} 2011, The Astrophysical
  Journal, 728, L26, \dodoi{10.1088/2041-8205/728/2/l26}

\bibitem[{{G{\'o}rski} {et~al.}(2005){G{\'o}rski}, {Hivon}, {Banday}, {Wand
  elt}, {Hansen}, {Reinecke}, \& {Bartelmann}}]{2005ApJ...622..759G}
{G{\'o}rski}, K.~M., {Hivon}, E., {Banday}, A.~J., {et~al.} 2005, \apj, 622,
  759, \dodoi{10.1086/427976}

\bibitem[{{Gunn} {et~al.}(2006){Gunn}, {Siegmund}, {Mannery}, {Owen}, {Hull},
  {Leger}, {Carey}, {Knapp}, {York}, {Boroski}, {Kent}, {Lupton}, {Rockosi},
  {Evans}, {Waddell}, {Anderson}, {Annis}, {Barentine}, {Bartoszek}, {Bastian},
  {Bracker}, {Brewington}, {Briegel}, {Brinkmann}, {Brown}, {Carr},
  {Czarapata}, {Drennan}, {Dombeck}, {Federwitz}, {Gillespie}, {Gonzales},
  {Hansen}, {Harvanek}, {Hayes}, {Jordan}, {Kinney}, {Klaene}, {Kleinman},
  {Kron}, {Kresinski}, {Lee}, {Limmongkol}, {Lindenmeyer}, {Long}, {Loomis},
  {McGehee}, {Mantsch}, {Neilsen}, {Neswold}, {Newman}, {Nitta}, {Peoples},
  {Pier}, {Prieto}, {Prosapio}, {Rivetta}, {Schneider}, {Snedden}, \&
  {Wang}}]{2006AJ....131.2332G}
{Gunn}, J.~E., {Siegmund}, W.~A., {Mannery}, E.~J., {et~al.} 2006, \aj, 131,
  2332, \dodoi{10.1086/500975}

\bibitem[{{Hauser} \& {Dwek}(2001)}]{2001ARA&A..39..249H}
{Hauser}, M.~G., \& {Dwek}, E. 2001, \araa, 39, 249,
  \dodoi{10.1146/annurev.astro.39.1.249}

\bibitem[{Hopkins {et~al.}(2007)Hopkins, Richards, \& Hernquist}]{Hopkins_2007}
Hopkins, P.~F., Richards, G.~T., \& Hernquist, L. 2007, The Astrophysical
  Journal, 654, 731, \dodoi{10.1086/509629}

\bibitem[{{Ikeda} {et~al.}(2011){Ikeda}, {Nagao}, {Matsuoka}, {Taniguchi},
  {Shioya}, {Trump}, {Capak}, {Comastri}, {Enoki}, {Ideue}, {Kakazu},
  {Koekemoer}, {Morokuma}, {Murayama}, {Saito}, {Salvato}, {Schinnerer},
  {Scoville}, \& {Silverman}}]{2011ApJ...728L..25I}
{Ikeda}, H., {Nagao}, T., {Matsuoka}, K., {et~al.} 2011, \apjl, 728, L25,
  \dodoi{10.1088/2041-8205/728/2/L25}

\bibitem[{{Jiang} {et~al.}(2015){Jiang}, {McGreer}, {Fan}, {Bian}, {Cai},
  {Cl{\'e}ment}, {Wang}, \& {Fan}}]{2015AJ....149..188J}
{Jiang}, L., {McGreer}, I.~D., {Fan}, X., {et~al.} 2015, \aj, 149, 188,
  \dodoi{10.1088/0004-6256/149/6/188}

\bibitem[{{Jiang} {et~al.}(2014){Jiang}, {Fan}, {Bian}, {McGreer}, {Strauss},
  {Annis}, {Buck}, {Green}, {Hodge}, {Myers}, {Rafiee}, \&
  {Richards}}]{2014ApJS..213...12J}
{Jiang}, L., {Fan}, X., {Bian}, F., {et~al.} 2014, \apjs, 213, 12,
  \dodoi{10.1088/0067-0049/213/1/12}

\bibitem[{{Jiang} {et~al.}(2016){Jiang}, {McGreer}, {Fan}, {Strauss},
  {Ba{\~n}ados}, {Becker}, {Bian}, {Farnsworth}, {Shen}, {Wang}, {Wang},
  {Wang}, {White}, {Wu}, {Wu}, {Yang}, \& {Yang}}]{2016ApJ...833..222J}
{Jiang}, L., {McGreer}, I.~D., {Fan}, X., {et~al.} 2016, \apj, 833, 222,
  \dodoi{10.3847/1538-4357/833/2/222}

\bibitem[{{Kim} \& {Im}(2021)}]{2021arXiv210306265K}
{Kim}, Y., \& {Im}, M. 2021, arXiv e-prints, arXiv:2103.06265.
\newblock \doarXiv{2103.06265}

\bibitem[{Kim {et~al.}(2020)Kim, Im, Jeon, Kim, Pak, Hyun, Taak, Shin, Lim,
  Paek, \& et~al.}]{Kim_2020}
Kim, Y., Im, M., Jeon, Y., {et~al.} 2020, The Astrophysical Journal, 904, 111,
  \dodoi{10.3847/1538-4357/abc0ea}

\bibitem[{Kirkpatrick {et~al.}(2011)Kirkpatrick, Schlegel, Ross, Myers,
  Hennawi, Sheldon, Schneider, \& Weaver}]{Kirkpatrick_2011}
Kirkpatrick, J.~A., Schlegel, D.~J., Ross, N.~P., {et~al.} 2011, The
  Astrophysical Journal, 743, 125, \dodoi{10.1088/0004-637x/743/2/125}

\bibitem[{{Kulkarni} {et~al.}(2019){Kulkarni}, {Worseck}, \&
  {Hennawi}}]{2019MNRAS.488.1035K}
{Kulkarni}, G., {Worseck}, G., \& {Hennawi}, J.~F. 2019, \mnras, 488, 1035,
  \dodoi{10.1093/mnras/stz1493}

\bibitem[{{La Franca} \& {Cristiani}(1997)}]{1997AJ....113.1517L}
{La Franca}, F., \& {Cristiani}, S. 1997, \aj, 113, 1517,
  \dodoi{10.1086/118369}

\bibitem[{{Lampton} {et~al.}(1976){Lampton}, {Margon}, \&
  {Bowyer}}]{1976ApJ...208..177L}
{Lampton}, M., {Margon}, B., \& {Bowyer}, S. 1976, \apj, 208, 177,
  \dodoi{10.1086/154592}

\bibitem[{{Lyke} {et~al.}(2020){Lyke}, {Higley}, {McLane}, {Schurhammer},
  {Myers}, {Ross}, {Dawson}, {Chabanier}, {Martini}, {Busca}, {Mas des
  Bourboux}, {Salvato}, {Streblyanska}, {Zarrouk}, {Burtin}, {Anderson},
  {Bautista}, {Bizyaev}, {Brandt}, {Brinkmann}, {Brownstein}, {Comparat},
  {Green}, {de la Macorra}, {Mu{\~n}oz Guti{\'e}rrez}, {Hou}, {Newman},
  {Palanque-Delabrouille}, {P{\^a}ris}, {Percival}, {Petitjean}, {Rich},
  {Rossi}, {Schneider}, {Smith}, {Vivek}, \& {Weaver}}]{2020ApJS..250....8L}
{Lyke}, B.~W., {Higley}, A.~N., {McLane}, J.~N., {et~al.} 2020, \apjs, 250, 8,
  \dodoi{10.3847/1538-4365/aba623}

\bibitem[{{Manti} {et~al.}(2017){Manti}, {Gallerani}, {Ferrara}, {Greig}, \&
  {Feruglio}}]{2017MNRAS.466.1160M}
{Manti}, S., {Gallerani}, S., {Ferrara}, A., {Greig}, B., \& {Feruglio}, C.
  2017, \mnras, 466, 1160, \dodoi{10.1093/mnras/stw3168}

\bibitem[{{Marshall} {et~al.}(1983){Marshall}, {Tananbaum}, {Avni}, \&
  {Zamorani}}]{1983ApJ...269...35M}
{Marshall}, H.~L., {Tananbaum}, H., {Avni}, Y., \& {Zamorani}, G. 1983, \apj,
  269, 35, \dodoi{10.1086/161016}

\bibitem[{{Masters} {et~al.}(2012){Masters}, {Capak}, {Salvato}, {Civano},
  {Mobasher}, {Siana}, {Hasinger}, {Impey}, {Nagao}, {Trump}, {Ikeda}, {Elvis},
  \& {Scoville}}]{2012ApJ...755..169M}
{Masters}, D., {Capak}, P., {Salvato}, M., {et~al.} 2012, \apj, 755, 169,
  \dodoi{10.1088/0004-637X/755/2/169}

\bibitem[{{Matsuoka} {et~al.}(2018){Matsuoka}, {Strauss}, {Kashikawa}, {Onoue},
  {Iwasawa}, {Tang}, {Lee}, {Imanishi}, {Nagao}, {Akiyama}, {Asami}, {Bosch},
  {Furusawa}, {Goto}, {Gunn}, {Harikane}, {Ikeda}, {Izumi}, {Kawaguchi},
  {Kato}, {Kikuta}, {Kohno}, {Komiyama}, {Lupton}, {Minezaki}, {Miyazaki},
  {Murayama}, {Niida}, {Nishizawa}, {Noboriguchi}, {Oguri}, {Ono}, {Ouchi},
  {Price}, {Sameshima}, {Schulze}, {Shirakata}, {Silverman}, {Sugiyama},
  {Tait}, {Takada}, {Takata}, {Tanaka}, {Toba}, {Utsumi}, {Wang}, \&
  {Yamashita}}]{2018ApJ...869..150M}
{Matsuoka}, Y., {Strauss}, M.~A., {Kashikawa}, N., {et~al.} 2018, \apj, 869,
  150, \dodoi{10.3847/1538-4357/aaee7a}

\bibitem[{McGreer {et~al.}(2018)McGreer, Fan, Jiang, \& Cai}]{McGreer_2018}
McGreer, I.~D., Fan, X., Jiang, L., \& Cai, Z. 2018, The Astronomical Journal,
  155, 131, \dodoi{10.3847/1538-3881/aaaab4}

\bibitem[{{McGreer} {et~al.}(2013){McGreer}, {Jiang}, {Fan}, {Richards},
  {Strauss}, {Ross}, {White}, {Shen}, {Schneider}, {Myers}, {Brandt}, {DeGraf},
  {Glikman}, {Ge}, \& {Streblyanska}}]{2013ApJ...768..105M}
{McGreer}, I.~D., {Jiang}, L., {Fan}, X., {et~al.} 2013, \apj, 768, 105,
  \dodoi{10.1088/0004-637X/768/2/105}

\bibitem[{{Niida} {et~al.}(2020){Niida}, {Nagao}, {Ikeda}, {Akiyama},
  {Matsuoka}, {He}, {Matsuoka}, {Toba}, {Onoue}, {Kobayashi}, {Taniguchi},
  {Furusawa}, {Harikane}, {Imanishi}, {Kashikawa}, {Kawaguchi}, {Komiyama},
  {Shirakata}, {Terashima}, \& {Ueda}}]{2020ApJ...904...89N}
{Niida}, M., {Nagao}, T., {Ikeda}, H., {et~al.} 2020, \apj, 904, 89,
  \dodoi{10.3847/1538-4357/abbe11}

\bibitem[{{Page} \& {Carrera}(2000)}]{2000MNRAS.311..433P}
{Page}, M.~J., \& {Carrera}, F.~J. 2000, \mnras, 311, 433,
  \dodoi{10.1046/j.1365-8711.2000.03105.x}

\bibitem[{{Palanque-Delabrouille} {et~al.}(2013){Palanque-Delabrouille},
  {Magneville}, {Y{\`e}che}, {Eftekharzadeh}, {Myers}, {Petitjean},
  {P{\^a}ris}, {Aubourg}, {McGreer}, {Fan}, {Dey}, {Schlegel}, {Bailey},
  {Bizayev}, {Bolton}, {Dawson}, {Ebelke}, {Ge}, {Malanushenko},
  {Malanushenko}, {Oravetz}, {Pan}, {Ross}, {Schneider}, {Sheldon}, {Simmons},
  {Tinker}, {White}, \& {Willmer}}]{2013AA...551A..29P}
{Palanque-Delabrouille}, N., {Magneville}, C., {Y{\`e}che}, C., {et~al.} 2013,
  \aap, 551, A29, \dodoi{10.1051/0004-6361/201220379}

\bibitem[{{Richards} {et~al.}(2002){Richards}, {Fan}, {Newberg}, {Strauss},
  {Vanden Berk}, {Schneider}, {Yanny}, {Boucher}, {Burles}, {Frieman}, {Gunn},
  {Hall}, {Ivezi{\'c}}, {Kent}, {Loveday}, {Lupton}, {Rockosi}, {Schlegel},
  {Stoughton}, {SubbaRao}, \& {York}}]{2002AJ....123.2945R}
{Richards}, G.~T., {Fan}, X., {Newberg}, H.~J., {et~al.} 2002, \aj, 123, 2945,
  \dodoi{10.1086/340187}

\bibitem[{Richards {et~al.}(2006)Richards, Strauss, Fan, Hall, Jester,
  Schneider, Berk, Stoughton, Anderson, Brunner, Gray, Gunn, Ivezi{\'{c}},
  Kirkland, Knapp, Loveday, Meiksin, Pope, Szalay, Thakar, Yanny, York,
  Barentine, Brewington, Brinkmann, Fukugita, Harvanek, Kent, Kleinman,
  Krzesi{\'{n}}ski, Long, Lupton, Nash, Eric H.~Neilsen, Nitta, Schlegel, \&
  Snedden}]{Richards_2006}
Richards, G.~T., Strauss, M.~A., Fan, X., {et~al.} 2006, The Astronomical
  Journal, 131, 2766, \dodoi{10.1086/503559}

\bibitem[{{Ross} {et~al.}(2012){Ross}, {Myers}, {Sheldon}, {Y{\`e}che},
  {Strauss}, {Bovy}, {Kirkpatrick}, {Richards}, {Aubourg}, {Blanton}, {Brandt},
  {Carithers}, {Croft}, {da Silva}, {Dawson}, {Eisenstein}, {Hennawi}, {Ho},
  {Hogg}, {Lee}, {Lundgren}, {McMahon}, {Miralda-Escud{\'e}},
  {Palanque-Delabrouille}, {P{\^a}ris}, {Petitjean}, {Pieri}, {Rich}, {Roe},
  {Schiminovich}, {Schlegel}, {Schneider}, {Slosar}, {Suzuki}, {Tinker},
  {Weinberg}, {Weyant}, {White}, \& {Wood-Vasey}}]{2012ApJS..199....3R}
{Ross}, N.~P., {Myers}, A.~D., {Sheldon}, E.~S., {et~al.} 2012, \apjs, 199, 3,
  \dodoi{10.1088/0067-0049/199/1/3}

\bibitem[{Ross {et~al.}(2013)Ross, McGreer, White, Richards, Myers,
  Palanque-Delabrouille, Strauss, Anderson, Shen, Brandt, Y{\`{e}}che, Swanson,
  Aubourg, Bailey, Bizyaev, Bovy, Brewington, Brinkmann, DeGraf, Matteo,
  Ebelke, Fan, Ge, Malanushenko, Malanushenko, Mandelbaum, Maraston, Muna,
  Oravetz, Pan, P{\^{a}}ris, Petitjean, Schawinski, Schlegel, Schneider,
  Silverman, Simmons, Snedden, Streblyanska, Suzuki, Weinberg, \&
  York}]{Ross_2013}
Ross, N.~P., McGreer, I.~D., White, M., {et~al.} 2013, The Astrophysical
  Journal, 773, 14, \dodoi{10.1088/0004-637x/773/1/14}

\bibitem[{Schindler {et~al.}(2019)Schindler, Fan, McGreer, Yang, Wang, Green,
  Fynbo, Krogager, Green, Huang, Kadowaki, Patej, Wu, \& Yue}]{Schindler_2019}
Schindler, J.-T., Fan, X., McGreer, I.~D., {et~al.} 2019, The Astrophysical
  Journal, 871, 258, \dodoi{10.3847/1538-4357/aaf86c}

\bibitem[{{Schmidt}(1963)}]{1963Natur.197.1040S}
{Schmidt}, M. 1963, \nat, 197, 1040, \dodoi{10.1038/1971040a0}

\bibitem[{{Schneider} {et~al.}(2001){Schneider}, {Fan}, {Strauss}, {Gunn},
  {Richards}, {Hill}, {MacQueen}, {Ramsey}, {Adams}, {Booth}, {Hill}, {Knapp},
  {Lupton}, {Saxe}, {Shetrone}, {Tufts}, {Vanden Berk}, {Wolf}, {York},
  {Anderson}, {Anderson}, {Bahcall}, {Brinkmann}, {Brunner}, {Csabai},
  {Fukugita}, {Hennessy}, {Ivezi{\'c}}, {Lamb}, {Munn}, \&
  {Thakar}}]{2001AJ....121.1232S}
{Schneider}, D.~P., {Fan}, X., {Strauss}, M.~A., {et~al.} 2001, \aj, 121, 1232,
  \dodoi{10.1086/319422}

\bibitem[{{Schneider} {et~al.}(2007){Schneider}, {Hall}, {Richards}, {Strauss},
  {Vanden Berk}, {Anderson}, {Brandt}, {Fan}, {Jester}, {Gray}, {Gunn},
  {SubbaRao}, {Thakar}, {Stoughton}, {Szalay}, {Yanny}, {York}, {Bahcall},
  {Barentine}, {Blanton}, {Brewington}, {Brinkmann}, {Brunner}, {Castander},
  {Csabai}, {Frieman}, {Fukugita}, {Harvanek}, {Hogg}, {Ivezi{\'c}}, {Kent},
  {Kleinman}, {Knapp}, {Kron}, {Krzesi{\'n}ski}, {Long}, {Lupton}, {Nitta},
  {Pier}, {Saxe}, {Shen}, {Snedden}, {Weinberg}, \& {Wu}}]{2007AJ....134..102S}
{Schneider}, D.~P., {Hall}, P.~B., {Richards}, G.~T., {et~al.} 2007, \aj, 134,
  102, \dodoi{10.1086/518474}

\bibitem[{{Schneider} {et~al.}(2010){Schneider}, {Richards}, {Hall}, {Strauss},
  {Anderson}, {Boroson}, {Ross}, {Shen}, {Brandt}, {Fan}, {Inada}, {Jester},
  {Knapp}, {Krawczyk}, {Thakar}, {Vanden Berk}, {Voges}, {Yanny}, {York},
  {Bahcall}, {Bizyaev}, {Blanton}, {Brewington}, {Brinkmann}, {Eisenstein},
  {Frieman}, {Fukugita}, {Gray}, {Gunn}, {Hibon}, {Ivezi{\'c}}, {Kent}, {Kron},
  {Lee}, {Lupton}, {Malanushenko}, {Malanushenko}, {Oravetz}, {Pan}, {Pier},
  {Price}, {Saxe}, {Schlegel}, {Simmons}, {Snedden}, {SubbaRao}, {Szalay}, \&
  {Weinberg}}]{2010AJ....139.2360S}
{Schneider}, D.~P., {Richards}, G.~T., {Hall}, P.~B., {et~al.} 2010, \aj, 139,
  2360, \dodoi{10.1088/0004-6256/139/6/2360}

\bibitem[{Shen {et~al.}(2020)Shen, Hopkins, Faucher-Giguère, Alexander,
  Richards, Ross, \& Hickox}]{10.1093/mnras/staa1381}
Shen, X., Hopkins, P.~F., Faucher-Giguère, C.-A., {et~al.} 2020, Monthly
  Notices of the Royal Astronomical Society, 495, 3252,
  \dodoi{10.1093/mnras/staa1381}

\bibitem[{{Shen} \& {Kelly}(2012)}]{2012ApJ...746..169S}
{Shen}, Y., \& {Kelly}, B.~C. 2012, \apj, 746, 169,
  \dodoi{10.1088/0004-637X/746/2/169}

\bibitem[{{Taylor}(2005)}]{2005ASPC..347...29T}
{Taylor}, M.~B. 2005, in Astronomical Society of the Pacific Conference Series,
  Vol. 347, Astronomical Data Analysis Software and Systems XIV, ed.
  P.~{Shopbell}, M.~{Britton}, \& R.~{Ebert}, 29

\bibitem[{{Vanden Berk} {et~al.}(2001){Vanden Berk}, {Richards}, {Bauer},
  {Strauss}, {Schneider}, {Heckman}, {York}, {Hall}, {Fan}, {Knapp},
  {Anderson}, {Annis}, {Bahcall}, {Bernardi}, {Briggs}, {Brinkmann}, {Brunner},
  {Burles}, {Carey}, {Castander}, {Connolly}, {Crocker}, {Csabai}, {Doi},
  {Finkbeiner}, {Friedman}, {Frieman}, {Fukugita}, {Gunn}, {Hennessy},
  {Ivezi{\'c}}, {Kent}, {Kunszt}, {Lamb}, {Leger}, {Long}, {Loveday}, {Lupton},
  {Meiksin}, {Merelli}, {Munn}, {Newberg}, {Newcomb}, {Nichol}, {Owen}, {Pier},
  {Pope}, {Rockosi}, {Schlegel}, {Siegmund}, {Smee}, {Snir}, {Stoughton},
  {Stubbs}, {SubbaRao}, {Szalay}, {Szokoly}, {Tremonti}, {Uomoto}, {Waddell},
  {Yanny}, \& {Zheng}}]{2001AJ....122..549V}
{Vanden Berk}, D.~E., {Richards}, G.~T., {Bauer}, A., {et~al.} 2001, \aj, 122,
  549, \dodoi{10.1086/321167}

\bibitem[{{Wang} {et~al.}(2016){Wang}, {Wu}, {Fan}, {Yang}, {Yi}, {Bian},
  {McGreer}, {Yang}, {Ai}, {Dong}, {Zuo}, {Jiang}, {Green}, {Wang}, {Cai},
  {Wang}, \& {Yue}}]{2016ApJ...819...24W}
{Wang}, F., {Wu}, X.-B., {Fan}, X., {et~al.} 2016, \apj, 819, 24,
  \dodoi{10.3847/0004-637X/819/1/24}

\bibitem[{Wang {et~al.}(2019)Wang, Yang, Fan, Wu, Yue, Li, Bian, Jiang,
  Ba{\~{n}}ados, Schindler, Findlay, Davies, Decarli, Farina, Green, Hennawi,
  Huang, Mazzuccheli, McGreer, Venemans, Walter, Dye, Lyke, Myers, \&
  Nunez}]{Wang_2019}
Wang, F., Yang, J., Fan, X., {et~al.} 2019, The Astrophysical Journal, 884, 30,
  \dodoi{10.3847/1538-4357/ab2be5}

\bibitem[{{Wang} {et~al.}(2021){Wang}, {Yang}, {Fan}, {Hennawi}, {Barth},
  {Banados}, {Bian}, {Boutsia}, {Connor}, {Davies}, {Decarli}, {Eilers},
  {Farina}, {Green}, {Jiang}, {Li}, {Mazzucchelli}, {Nanni}, {Schindler},
  {Venemans}, {Walter}, {Wu}, \& {Yue}}]{2021ApJ...907L...1W}
{Wang}, F., {Yang}, J., {Fan}, X., {et~al.} 2021, \apjl, 907, L1,
  \dodoi{10.3847/2041-8213/abd8c6}

\bibitem[{{Wolf} {et~al.}(2003){Wolf}, {Wisotzki}, {Borch}, {Dye},
  {Kleinheinrich}, \& {Meisenheimer}}]{2003AA...408..499W}
{Wolf}, C., {Wisotzki}, L., {Borch}, A., {et~al.} 2003, \aap, 408, 499,
  \dodoi{10.1051/0004-6361:20030990}

\bibitem[{{Worseck} {et~al.}(2011){Worseck}, {Prochaska}, {McQuinn},
  {Dall'Aglio}, {Fechner}, {Hennawi}, {Reimers}, {Richter}, \&
  {Wisotzki}}]{2011ApJ...733L..24W}
{Worseck}, G., {Prochaska}, J.~X., {McQuinn}, M., {et~al.} 2011, \apjl, 733,
  L24, \dodoi{10.1088/2041-8205/733/2/L24}

\bibitem[{{Yang} {et~al.}(2016){Yang}, {Wang}, {Wu}, {Fan}, {McGreer}, {Bian},
  {Yi}, {Yang}, {Ai}, {Dong}, {Zuo}, {Green}, {Jiang}, {Wang}, {Wang}, \&
  {Yue}}]{2016ApJ...829...33Y}
{Yang}, J., {Wang}, F., {Wu}, X.-B., {et~al.} 2016, \apj, 829, 33,
  \dodoi{10.3847/0004-637X/829/1/33}

\bibitem[{{Y{\`e}che} {et~al.}(2010){Y{\`e}che}, {Petitjean}, {Rich},
  {Aubourg}, {Busca}, {Hamilton}, {Le Goff}, {Paris}, {Peirani}, {Pichon},
  {Rollinde}, \& {Vargas-Maga{\~n}a}}]{2010A&A...523A..14Y}
{Y{\`e}che}, C., {Petitjean}, P., {Rich}, J., {et~al.} 2010, \aap, 523, A14,
  \dodoi{10.1051/0004-6361/200913508}

\bibitem[{{Yuan} {et~al.}(2013){Yuan}, {Zhang}, {Zhang}, {Lei}, {Dong}, \&
  {Zhao}}]{2013A&C.....3...65Y}
{Yuan}, H., {Zhang}, H., {Zhang}, Y., {et~al.} 2013, Astronomy and Computing,
  3, 65, \dodoi{10.1016/j.ascom.2013.12.001}

\bibitem[{Zhan(2021)}]{Zhan_2021}
Zhan, H. 2021, Chinese Science Bulletin, 66, 1290, \dodoi{10.1360/tb-2021-0016}

\end{thebibliography}
\bibliographystyle{aasjournal1}

\end{document}